\documentclass[
aps,
prb,
reprint,
superscriptaddress,
amsmath,
amssymb,
longbibliography,
]{revtex4-1}

\usepackage{graphicx}
\usepackage{dcolumn}
\usepackage{bm}
\usepackage[version=3]{mhchem}
\usepackage{natbib,hyperref}
\hypersetup{colorlinks=true,linkcolor=blue,citecolor=blue,urlcolor=blue}
\usepackage{color}
\usepackage{float}

\renewcommand{\vec}[1]{\mathbf{#1}}

\begin{document}


\title{Exciton spin dynamics in GaSe}

\author{Yanhao Tang}
\author{Wei Xie}
\affiliation{Department of Physics and Astronomy, Michigan State University, East Lansing, MI 48824, USA}
\author{Krishna C. Mandal}
\affiliation{Department of Electrical Engineering, University of South Carolina, Columbia, SC 29208, USA}
\author{John A. McGuire}
\email{mcguire@pa.msu.edu}
\author{Chih Wei Lai}
\email{cwlai@msu.edu}
\affiliation{Department of Physics and Astronomy, Michigan State University, East Lansing, MI 48824, USA}



\begin{abstract}
We analyze exciton spin dynamics in GaSe under \emph{nonresonant} circularly polarized optical pumping with an exciton spin-flip rate-equation model. The model reproduces polarized time-dependent photoluminescence measurements in which the initial circular polarization approaches unity even when pumping with 0.15 eV excess energy. At T = 10 K, the exciton spin relaxation exhibits a biexponential decay with a sub-20 ps and a $>$500 ps time constants, which are also reproduced by the rate-equation model assuming distinct spin-relaxation rates for \emph{hot} (nonequilibrium) and \emph{cold} band-edge excitons.
\end{abstract}


\maketitle

\section{Introduction}
High spin polarization and long spin-relaxation times in solid-state systems are desirable for many spintronic applications. Toward this goal, nonequilibrium spin dynamics have been investigated in various semiconductors \cite{dyakonov1984,dyakonov2008,pikus1984,wu2010,zutic2004,awschalom2013}, with gallium arsenide (GaAs) the most studied example. However, the degenerate heavy- and light-hole valence bands and sub-ps hole-spin relaxation in bulk GaAs have limited optically pumped electron-spin polarization in GaAs to 1/2 and the degree of circular polarization of the resulting photoluminescence (PL) to 1/4 \cite{dyakonov1984,dyakonov2008}. The electron-spin relaxation times can be enhanced by doping \cite{kikkawa1997} or quantum confinement \cite{ohno1999}. For example, near unity electron spin polarization can be obtained in heterostructures where heavy- and light-hole energy degeneracy is lifted by quantum confinement or strain \cite{kohl1991,amand1994,pfalz2005}.

In a recent report \cite{tang2015}, we demonstrated that generation and preservation of a high degree of optical spin polarization, $\rho$, is possible in nanoscale slabs of GaSe even when optically pumping 0.1 to 0.2 eV above the gap. This is a consequence of the unique bandstructure (Fig.~\ref{fig:band}) of the group-III monochalcogenides \cite{gamarts1977,ivchenko1977}, which consists of orbitally nondegenerate uppermost valence band (UVB) and lowermost conduction band (LCB) with minimal spin--orbit-induced spin splittings \cite{do2015}. Most importantly, in GaSe, the large separation of the LCB and UVB from adjacent bands results in the suppression of Elliot-Yafet spin relaxation mechanism. In this study, we show that the exciton spin dynamics in GaSe at low temperature can be quantitatively reproduced by a simple model including momentum relaxation of excitons and spin relaxation of excitons and carriers.


\section{Crystal and Band Structures}\label{sec:bandstructure}
The III-VI semiconducting compounds GaS, GaSe, and GaTe (MX) all form layered crystals. The bonding between the layers is weak, resulting in the easy cleavage of these crystals. In the case of GaS and GaSe, each layer consists of four planes of atoms in the sequence X-M-M-X and belong to the space group $D^1_{3h}$. Different stacking orders of hexagonal layers result in formation of four commonly known polytypes: $\epsilon (D^1_{3h})$, $\beta$ ($D^4_{6h}$), rhombohedral $\gamma (C^5_{3v})$, and $\delta (C^4_{6v})$. Symmetry-dependent optical transitions and spin dynamics near the band edge can be sensitive to polytypes \cite{brebner1967}. Here, we study $\epsilon$-GaSe which has an ABA (Bernal) stacking order and belongs to space group $D_{3h}^1-P\bar{6}m2$, which is noncentrosymmetric, i.e., lacking a spatial inversion center.

\begin{figure}[htb!]
\includegraphics[width=0.48 \textwidth]{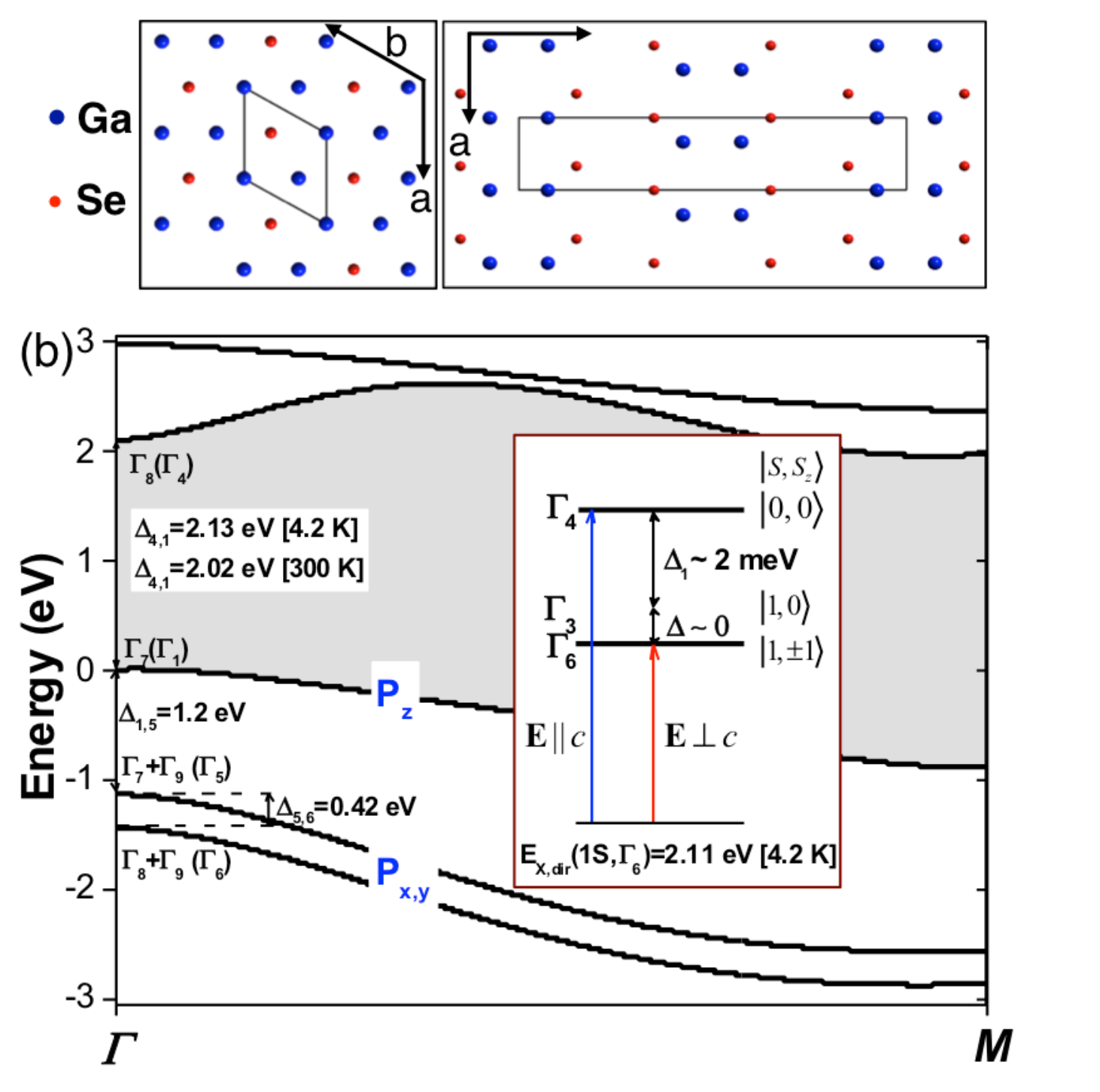}
\caption{\label{fig:band}\textbf{Band structure and selection rules.} (a) Crystal structure of $\epsilon$-GaSe showing $ABA$ stacking of the individual layers, space group $D_{3h}^{1}/P\bar{6}m2$ (\#187) and point group $D_{3h}$. An individual layer consists of four planes of \ce{Se-Ga-Ga-Se}, with the \ce{Ga-Ga} bond normal to the layer plane arranged on a hexagonal lattice and the Se anions located in the eclipsed conformation when viewed along the $c$-axis. The inner solid boxes represent a unit cell. (b) Sketches of the band structure of $\epsilon$-GaSe at the $\Gamma$-point and the representations to which the states at the $\Gamma$-point belong with (without) spin--orbit interaction. (inset) Direct-gap excitons and selection rules.}
\end{figure}

Bulk GaSe is generally regarded as an indirect-band-gap semiconductor. An indirect transition from the $\Gamma$-point to the $M$-point $\sim$10-20 meV below the direct gap at the $\Gamma$-point is nearly resonant with the direct exciton transitions at the $\Gamma$-point (exciton binding energy 20-30 meV) \cite{brebner1967,aulich1969,mercier1973,mooser1973,schluter1976,le-toullec1977,le-toullec1980,sasaki1981,capozzi1993}. Fig.~\ref{fig:band} shows a schematic crystal structure and bandstructure of bulk $\epsilon$-GaSe. The LCB and UVB near the $\Gamma$-point are derived primarily from Ga $s$-like and Se $p_z$-like orbitals and have respectively $\Gamma_4$ and $\Gamma_1$ symmetry. The valence bands are split by crystal-field anisotropy and spin--orbit interaction, leading to two bands with Se $p_x, p_y$ symmetry about 1.2 and 1.6 eV below the UVB maximum.

The strongly anisotropic crystal structure leads to correspondingly anisotropic optical properties. The polarization vectors $\vec{E} \parallel c$ and $\vec{E} \perp c$ belong to the $\Gamma_{4}$ and $\Gamma_{6}$ representations, respectively. Considering the transition between UVB ($\Gamma_1$) and LCB ($\Gamma_4$) without spin, the direct product $\Gamma_1 \times \Gamma_4$ belongs to the representation $\Gamma_4$ of $D_{3h}$; therefore, the direct transition $\Gamma_1 \rightarrow \Gamma_4$ is orbitally allowed only for $\vec{E} \parallel c$. Taking into account spin, the direct transitions then occur between valence and conduction bands with the following symmetries in the double group $\bar{D}_{3h}$:

\begin{align}
\Gamma_4 &\rightarrow \Gamma_8 \, (s\text{-like conduction band}) \nonumber \\
\Gamma_1 &\rightarrow \Gamma_7 \, (B, p_z\text{-like valence band}) \nonumber \\
\Gamma_5 &\rightarrow \Gamma_7 + \Gamma_9 \, (A, p_{x,y}\text{-like valence band}) \nonumber \\
\Gamma_6 &\rightarrow \Gamma_8 + \Gamma_9 \, (C, p_{x,y}\text{-like valence band})  
\end{align}

\noindent The direct product $\Gamma_7 \times \Gamma_8 = \Gamma_3 + \Gamma_4 + \Gamma_6$ then contains the representations for both $\vec{E} \perp c$ and $\vec{E} \parallel c$ so that the optical transition between the UVB and LCB becomes weakly dipole-allowed for $\vec{E} \perp c$.

A complete accounting of the optical response requires an excitonic (two-particle) picture. To a first approximation, one can regard band-edge excitons derived from carriers only in the conduction band and the $\Gamma_1$ UVB. This structure is then perturbed by spin--orbit mixing of the $\Gamma_5$ valence band into the UVB and then further perturbed by electron-hole exchange. The direct transition from the electronic ground state to an electron ($\Gamma_8$) and hole ($\Gamma_7$) in an $s$-like ($\Gamma_1$) direct-gap exciton state yields an exciton belonging to the following (degenerate) representations \cite{mooser1973,schluter1976,gamarts1977,ivchenko1977,sasaki1981}:

\begin{equation}
\Gamma^{(s)}_{X}=\Gamma_7 \times \Gamma_8 \times \Gamma_1 = \Gamma_4 + \Gamma_3 + \Gamma_6,
\end{equation}
where $\Gamma_4$ is a pure spin singlet and $\Gamma_3$ and $\Gamma_6$ are pure spin triplet states. 

Including spin--orbit mixing of the $\Gamma_5$ lower valence band into the UVB, the $s$-type band-edge excitonic wavefunctions become \cite{mooser1973}
\begin{align}
\Gamma_4 &= \left| S \right> + \alpha_4 \left| T \right>, \nonumber \\
\Gamma_3 &= \left| T \right>, \nonumber \\
\Gamma_6 &= \left| T \right> + \alpha_6 \left| S \right>,
\end{align}
where $\left| S \right>$ and $\left| T \right>$ represent singlet and triplet states, respectively. Using $\Delta_{SO}$ measured by Sasaki et al. \cite{sasaki1975a,sasaki1975b}, the coefficients $\alpha_4 \approx \alpha_6 \approx (\Delta_{SO}/E_{BA})\approx(0.44\,\textrm{eV}/1.27\,\textrm{eV}) \approx 0.35$. The $\Gamma_5$ state is mixed into the UVB at about the 10\% level, consistent with the ratio of the oscillator strength (absorbance) of $\vec{E} \perp c$ and $\vec{E} \parallel c$. The upper level $\Gamma_4$ corresponds to a total exciton spin $S$ = 0 and experiences a splitting $\Delta_1 \approx 2$ meV because of electron--hole exchange \cite{mooser1973}. The states $\Gamma_3$ and $\Gamma_6$ correspond to a total exciton spin $S$ = 1, and $S_z$= 0, $\pm1$ and are nearly degenerate (energy splitting $\Delta \, \approx \, 0$). These states are thus labeled by the indices 0 and $\pm1$. The $\Gamma_4$ state can be excited by light with $\vec{E} \, \parallel \, c$. For optical excitation with wave vector $\vec{k} \parallel c$, $\Gamma_6$ ($S_z \, = \pm1$) states can be excited by circularly polarized light with $\vec{E} \perp c$, whereas the $\Gamma_3$ state is optically inactive \cite{gamarts1977,ivchenko1977}.

Earlier studies suggested that a high degree of optical orientation in GaSe could be achieved under nearly resonant excitation at low temperatures. Using steady-state measurements, Gamarts et al. \cite{gamarts1977,ivchenko1977} demonstrated optical orientation and alignment of excitons in GaSe by showing luminescence with circular polarization above 90\% under steady-state circularly polarized optical excitation \emph{in resonance} with direct excitons at cryogenic temperatures.

\section{Results}

Sample preparation and experimental measurement methods were described previously \cite{tang2015}. Thin films of GaSe crystals were mechanically exfoliated from a Bridgman-grown crystal \cite{mandal2008a} and deposited onto a silicon substrate with a 90 nm \ce{SiO2} layer, with the thickness measured using atomic force microscopy. Samples were mounted in vacuum on a copper cold finger attached to an optical liquid helium flow cryostat for all experiments. GaSe nanoslabs were optically excited by 2 ps laser pulses from a synchronously pumped optical parametric oscillator ($\lambda_p \sim$ 560--595 nm, $E_p \sim$ 2.21--2.08 eV) or by second-harmonic pulses from a Ti:sapphire oscillator ($\lambda_p \sim$ 410 nm, $E_p \sim$ 3.0 eV). The laser beam was focused through a microscope objective (numerical aperture N.A. = 0.28) to an area of about 80 $\mu$m$^2$ on the sample. The wave vector of the pump is along the crystal $c$-axis (the surface normal), i.e., the electric field vector $\vec{E}$ is orthogonal to the $c$-axis ($\vec{E} \perp c$). The polarization and the flux ($P$) of the pump laser were controlled by liquid-crystal-based devices without mechanical moving parts. The samples were excited with the pump flux $P$ from $0.1$ $P_0$ to $P_0$, where $P_0 = 2.6\times10^{14}$~cm$^{-2}$ photons per pulse. We estimate the photoexcitation density to be from $\approx 2\times10^{16}$ cm$^{-3}$ to $3.4\times10^{17}$ cm$^{-3}$ ($2.7\times10^{-10}$ cm$^{-2}$ per layer) considering the absorption coefficient at 2.1 eV ($\approx10^3$ cm$^{-1}$ for $\vec{E} \perp c$) and Fresnel loss from reflection. The photoexcited carrier density is below the Mott transition \cite{pavesi1989} of direct excitons occurring near electron-hole (\emph{e-h}) pair densities of $4\times10^{17}$ cm$^{-3}$.

\begin{figure}[htb]
\centering
\includegraphics[width=0.45 \textwidth]{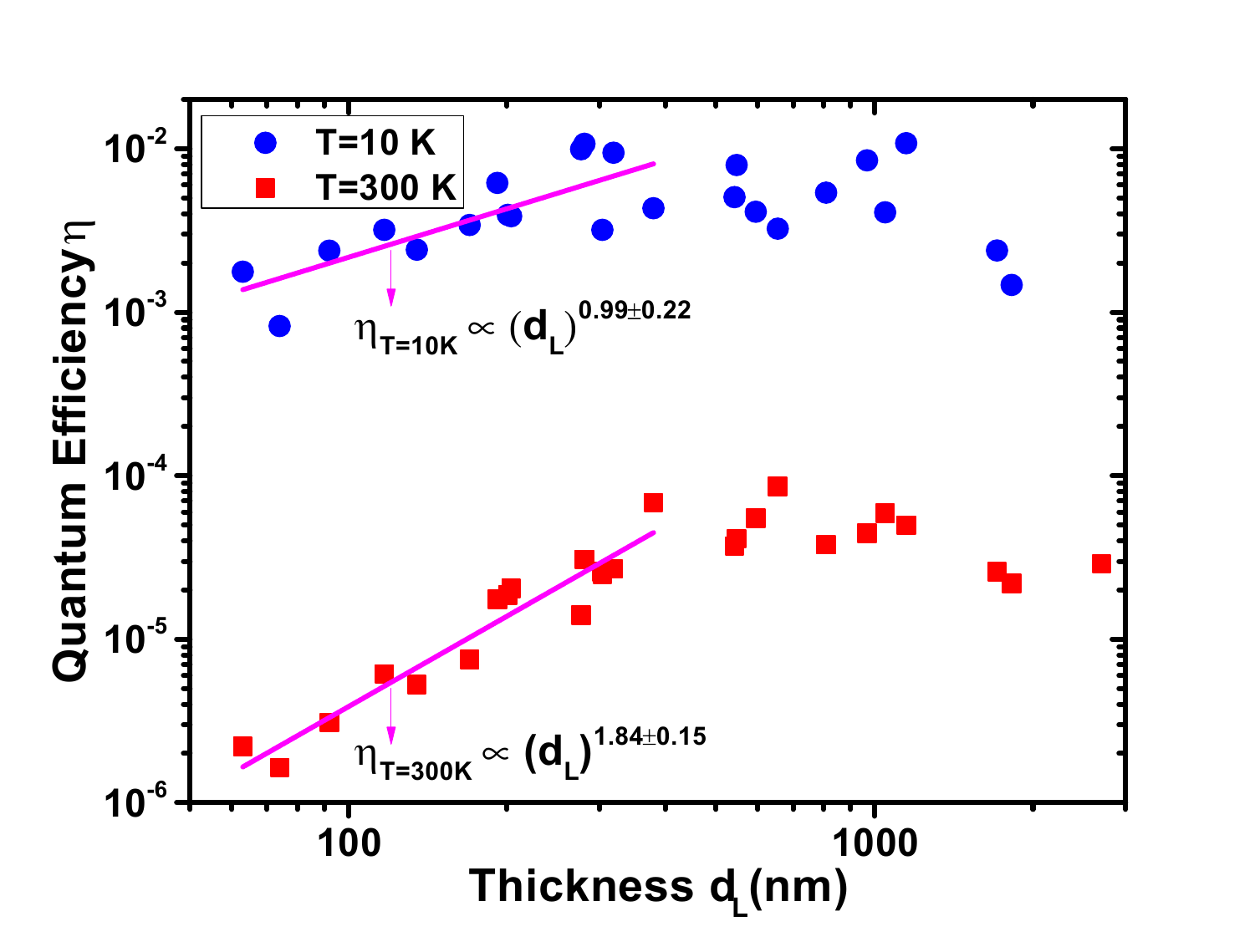}
\caption{\label{fig:qe}\textbf{PL quantum efficiency.} Quantum efficiency of luminescence as a function of thickness of GaSe nanoslabs at T = 300 and 10 K. Here, quantum efficiency is defined as the ratio between the luminescence emission flux and optical absorption flux per layer. Optical absorption flux is determined using experimentally measured reflectance for each sample and reported absorption coefficient $\alpha= 1.1\times10^3$ cm$^{-1}$ \cite{le-toullec1977,*le-toullec1980,*piccioli1977,*adachi1992}. Optical collection and detection efficiency is measured by passing a 633-nm laser beam with known power through the optical set-up and spectrometer. The emission flux is then calculated by including Fresnel reflection loss at the surface assuming angularly isotropic emission.}
\end{figure}

Time- and polarization-resolved PL measurements allow us to determine separately the recombination time, the initial spin orientation, and the spin-relaxation time. Polarized PL measurements were performed under excitation with excess energy of about 0.1 to 0.2 eV above the exciton emission peak. The band-edge exciton PL emission at room temperature is near 620 nm (2.0 eV), independent of thickness. The exciton peak gradually red shifts from 590 nm (2.1 eV) in bulk (thickness$>1000$ nm) to 620 nm (2.0 eV) to 90-nm nanoslabs at T = 10 K. This shift is attributed to increasing contribution to PL from localized excitons, which is the subject of ongoing studies and beyond the scope of this paper. In addition, we observe that the quantum yield of luminescence is greatly suppressed in sub-50-nm thick samples (Fig.~\ref{fig:qe}). In this paper, we focus on a 540 nm thick GaSe sample (S540nm). 

\begin{figure}[htb]
\centering
\includegraphics[width=0.45 \textwidth]{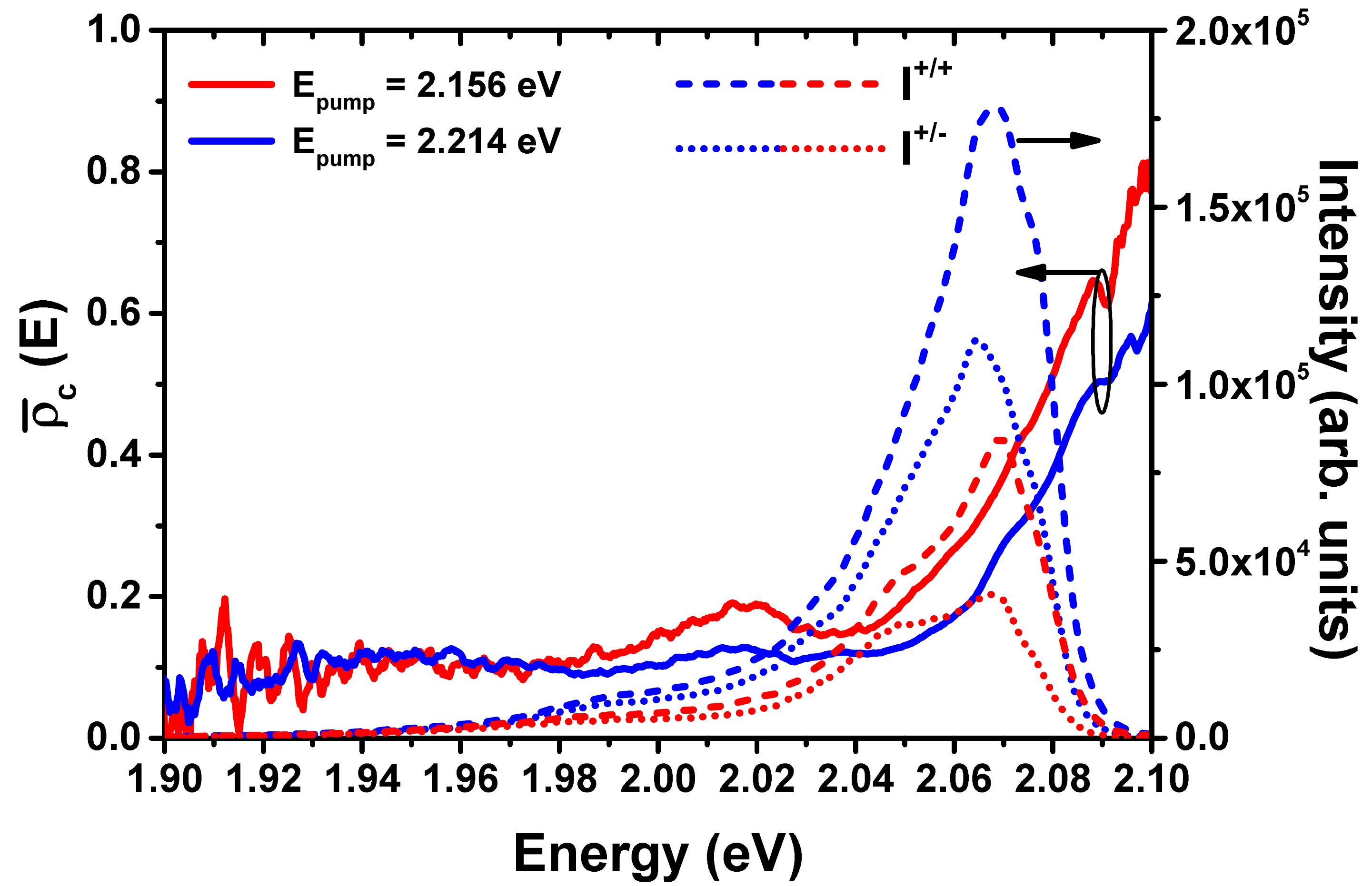}
\caption{\textbf{Polarized PL spectra under optical excitation at 2.156 eV and 2.214 eV.} Time-integrated PL spectra [$I^{+/+}(E)$(co-circular, dashed lines) and $I^{+/-}(E)$(cross-circular, dotted lines)] and degree of circular polarization $\bar{\rho}_c(E)$ (solid lines) of S540nm GaSe samples under $\sigma^+$ excitation at pump flux $P$ = 0.5 $P_0$, where $P_0$ = $2.6\times10^{14}$  cm$^{-2}$ per pulse. Blue (Red) lines are for $E_{pump}$ = 2.214 eV (2.156 eV). 
}\label{fig:ex560-570nm_spec}
\end{figure}

\begin{figure}[htb]
\centering
\includegraphics[width=0.45 \textwidth]{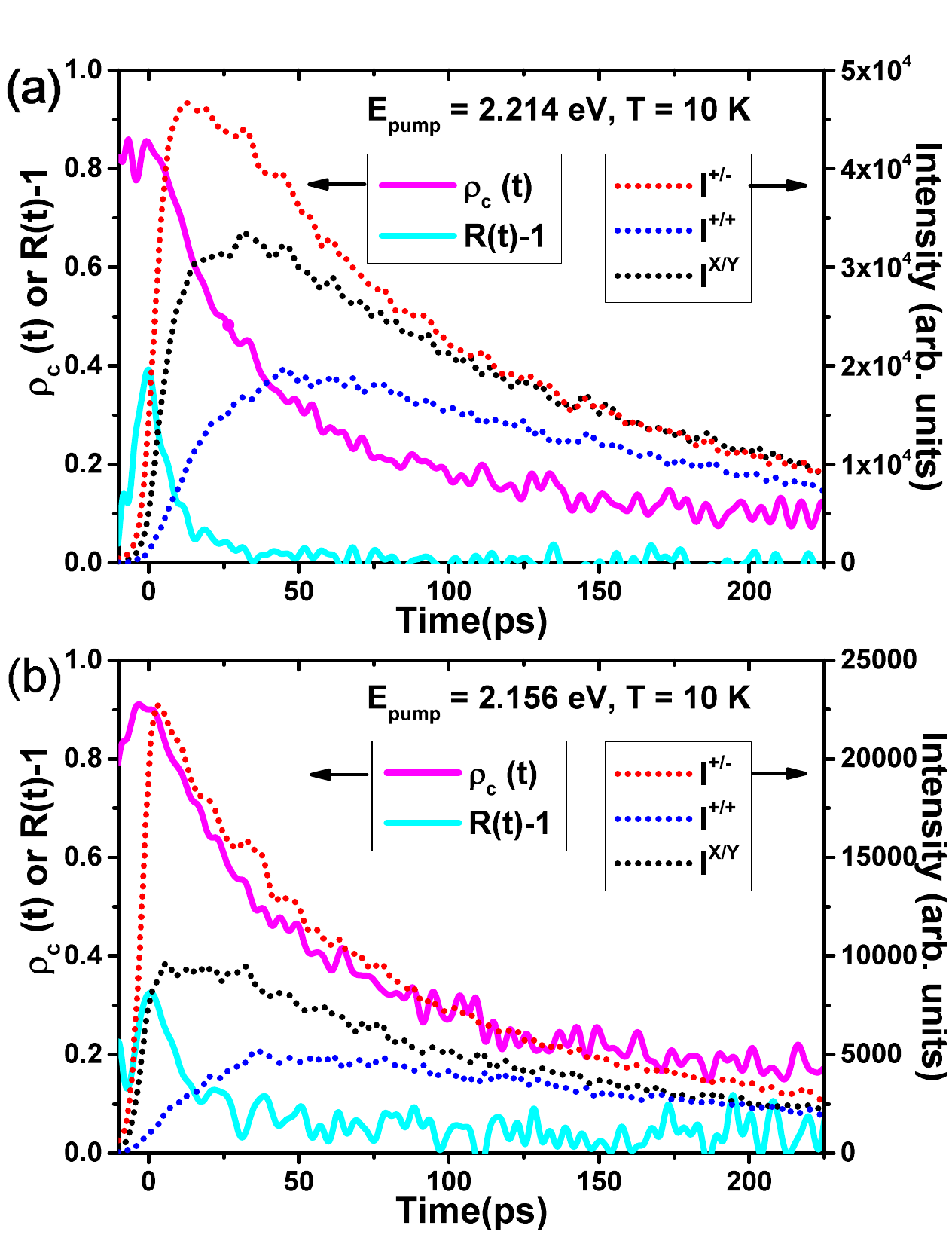}
\caption{\textbf{Polarized PL dynamics under optical excitation at 2.156 eV and 2.214 eV.} (a) Time-dependent PL intensity [$I^{+/+}(t)$, $I^{+/-}(t)$, and $I^{X/Y}(t)$ (dotted red, blue, and black, respectively)], degree of circular polarization $\rho_c(t)=\frac{I^{+/+}(t)-I^{+/-}(t)}{I^{+/+}(t)+I^{+/-}(t)}$ (solid magenta), and $R(t)-1= \frac{I^{+/+}(t)+I^{+/-}(t)}{I^{X/X}(t)+I^{X/Y}(t)}-1$ (solid cyan) under excitation $E_{pump}$ = 2.214 eV at $P$ = 0.5 $P_0$. (b) Same as (a), but for excitation $E_{pump}$ = 2.156 eV. The $\pm/\pm$ ($X/Y$) labels represent the helicity (polarization) of the pump laser and luminescence, respectively.
}\label{fig:ex560-570nm_dyn}
\end{figure}

Fig. \ref{fig:ex560-570nm_spec} shows time-integrated PL spectra under co- and cross-circularly polarized excitation and detection at T = 10 K. The PL is peaked at about 2.07 eV with a low-energy tail that is attributed to localized excitons. As in samples described previously \cite{tang2015}, the degree of steady-state spin polarization ($\bar{\rho}_c$) is highest at high energies, as expected given the reduced time spent at such energies. There is also a pronounced increase in $\bar{\rho}_c$ as the excitation photon energy decreases towards the band gap. This is as expected from the greater spin--orbit field at larger $k$ in the D'yakonov--Perel' (DP) mechanism of spin relaxation and the increased rate of scattering at larger $k$ in the Elliott--Yafet (EY) mechanism.

For more direct insights into the nature of the spin relaxation, we performed time-resolved measurements of the spectrally integrated PL (Fig. \ref{fig:ex560-570nm_dyn}). The PL at cryogenic temperatures (10 K) reveals a fast ($<$10 ps) rise time followed by a biexponential decay with time constants $\tau'_0\approx$ 20--50~ps and $\tau''_0\approx$ 150--200~ps in all samples. The time-dependent circular polarization $\rho_c(t)$ is also observed to be biexponential with decay time constants $\tau'_s \approx$ 30--40~ps and $\tau''_s \gtrsim$ 150--200~ps. The initial, resolution-limited rise in the PL signal represents the rate of carrier scattering to small values of the momentum (i.e., to near the band edge at the $\Gamma$-point). The biexponential nature of $\rho_c(t)$ can be understood as a consequence of fast spin relaxation during thermalization to the band edge followed by slower spin relaxation once the carriers are at the band edge. The initial decay of spin polarization is greater at 2.214 eV than at 2.156 eV, again consistent with the expectation that greater spin--orbit field and faster momentum scattering at higher energies should result in faster spin relaxation.

Previously, we showed that carrier spin relaxation leads to rotation of the emission dipoles from in-plane to out-of-plane, i.e., from perpendicular to the $c$-axis to along the $c$-axis \cite{tang2015a}. This is manifested by the polarized remote-edge luminescence (REL) 100 $\mu$m away from the pumping spot at the cleaved edges of the sample as a result of index-guided (waveguided) emission from out-of-plane dipoles \cite{tang2015a}. Time-resolved measurements of the spectrally integrated REL at $T=10$~K reveal an initial rise of the REL with a $\sim$30~ps timescale that matches the initial fast decay of the PL from the excitation spot.

\section{Spin relaxation in G\lowercase{a}S\lowercase{e}}
Spin-relaxation mechanisms in semiconductors depend on details of the energy versus momentum dispersion and spin-splitting because of spin--orbit interaction \cite{luttinger1955,kane1957}. In III--V semiconductors such as GaAs, the group-V $p_x$, $p_y$, and $p_z$-derived valence bands are closely spaced because of the small crystal field. Consequently, the angular momentum and quasimomentum of holes are strongly coupled, and the spin orientation of holes is lost in a period comparable with the momentum relaxation time ($\tau_p$) \cite{fishman1977,zerrouati1988,hilton2002,yu2005,shen2010,dyakonov2008,amand2008,wu2010}. The slower relaxation of \emph{electron} spin in such semiconductors is a result of the reduced spin--orbit mixing of the group-III $s$-orbital-derived conduction band with distant bands. Electron-spin relaxation is usually analyzed in terms of three mechanisms \cite{dyakonov1984,pikus1984,wu2010,boross2013}: D'yakonov--Perel' (DP) \cite{dyakonov1971,dyakonov1972,dyakonov1984}, which is associated with spin--orbit-induced spin splitting of a band in noncentrosymmetric systems \cite{dyakonov1972}; Elliot--Yafet (EY) \cite{elliott1954,yafet1963}, which is associated with spin--orbit-induced band-mixing; and Bir--Aronov--Pikus (BAP) \cite{bir1973,pikus1974,bir1975,aronov1983}, which is associated with electron--hole exchange interactions. For photoexcited carrier densities above $10^{16}$ cm$^{-3}$ as studied in undoped GaSe here, we neglect spin relaxation due to the BAP mechanism.


The EY and DP mechanisms are associated with the spin--orbit interaction (SOI) and the spin--orbit-induced spin splitting, $\Delta_s (\vec{k}) = |E(\vec{k}, \uparrow) - E(\vec{k}, \downarrow)|$ \cite{dyakonov2008,boross2013}. Using a four-state (two bands with spin) model Hamiltonian in the absence of an external magnetic field, one can relate the spin-relaxation rate of electrons (holes) with quasi-wave vector $\vec{k}$ away from the conduction (valence) band edge with the following equation \cite{boross2013}: 
\begin{equation}
	\Gamma_s \sim \frac{|\Delta_s(\vec{k})|^2}{\Gamma_p} + \frac{\Gamma_p |L(\vec{k})|^2}{\Gamma_p^2+\Delta_g^2(\vec{k})},
\end{equation}
where $\Gamma_p = \hbar/\tau_p$ is the scattering rate of electron/hole, with $\tau_p$ being the corresponding momentum scattering (or correlation) time, $\Delta_s(\vec{k})$ being the spin--orbit-induced spin splitting, and $L(\vec{k})$ being the SOI between the adjacent bands with energy separation $\Delta_g$. 

In GaSe, the initial carrier cooling to the band edge occurs in the sub-ps to sub-10 ps range (Fig. \ref{fig:ex560-570nm_dyn}) \cite{nusse1997}. This fast energy and momentum relaxation ($\tau_p \sim$ 1 ps) is demonstrated by the sub-10 ps PL rise time that is nearly independent of PL emission energy. 
The $p_z$-like UVB is well isolated from the LVB and the adjacent $p_{x,y}$-like valence bands (i.e., $\Delta_g\sim$1--2 eV), and as a result $L/\Delta_g \approx$ 0.02--0.04 \cite{kuroda1980,kuroda1981}. The hole-spin relaxation caused by the EY mechanism $\Gamma_s^{EY} \approx (\frac{L}{\Delta_g})^2 \Gamma_p$ ($\Gamma_p \ll \Delta_g(\vec{k})$) is expected to be much smaller than the momentum relaxation rate $\Gamma_p$, resulting in a spin-relaxation time of $\sim$1000 ps or more for $\tau_p\sim$1 ps. 

In $\epsilon$-GaSe, hole spin splitting $\Delta^h_s(k'=0.15) \approx$ 5 meV \cite{do2015}, where $k' = |\vec{k}| / \overline{\Gamma K}$. As a result, for holes with finite momentum $k'$, $|\Delta^h_s(k')|^2/\Gamma_p \gg \Gamma_p |L(k')|^2/\Delta_g^2(k')$, i.e., spin relaxation of \emph{hot} holes in GaSe is dominated by the DP mechanism, in which spin relaxation occurs from the precession of spins in an effective magnetic field associated with $\Delta^h_s(k')$ \cite{dyakonov1984,dyakonov2008,boross2013}. When the DP mechanism is dominant, the smaller the spin splitting, the longer the spin relaxation time $\tau_s \approx \hbar/\Gamma_s$ for the same momentum relaxation rate $\Gamma_p$. 

The momentum scattering time $\tau_p(n)$ is expected to decrease with increasing carrier density $n$. Therefore, the spin relaxation time, $\tau_s$, should increase with increasing $n$ if spin relaxation is dominated by the DP mechanism and decrease with increasing $n$ if spin relaxation is dominated by the EY mechanism. In our experiments, at T= 10 K, we found the initial decay of $\rho_c$ is characterized by $\tau'_s\propto n^{-0.23}$. The decreasing spin relaxation time with the increasing density (i.e., decreasing $\tau_p$) suggests that the EY mechanism plays a larger role than the DP mechanism in the spin relaxation of \emph{cold} excitons near the band edge in GaSe at low temperature.

\section{Theoretical Modeling}
\begin{figure}[htb]
\centering
\includegraphics[width=0.35 \textwidth]{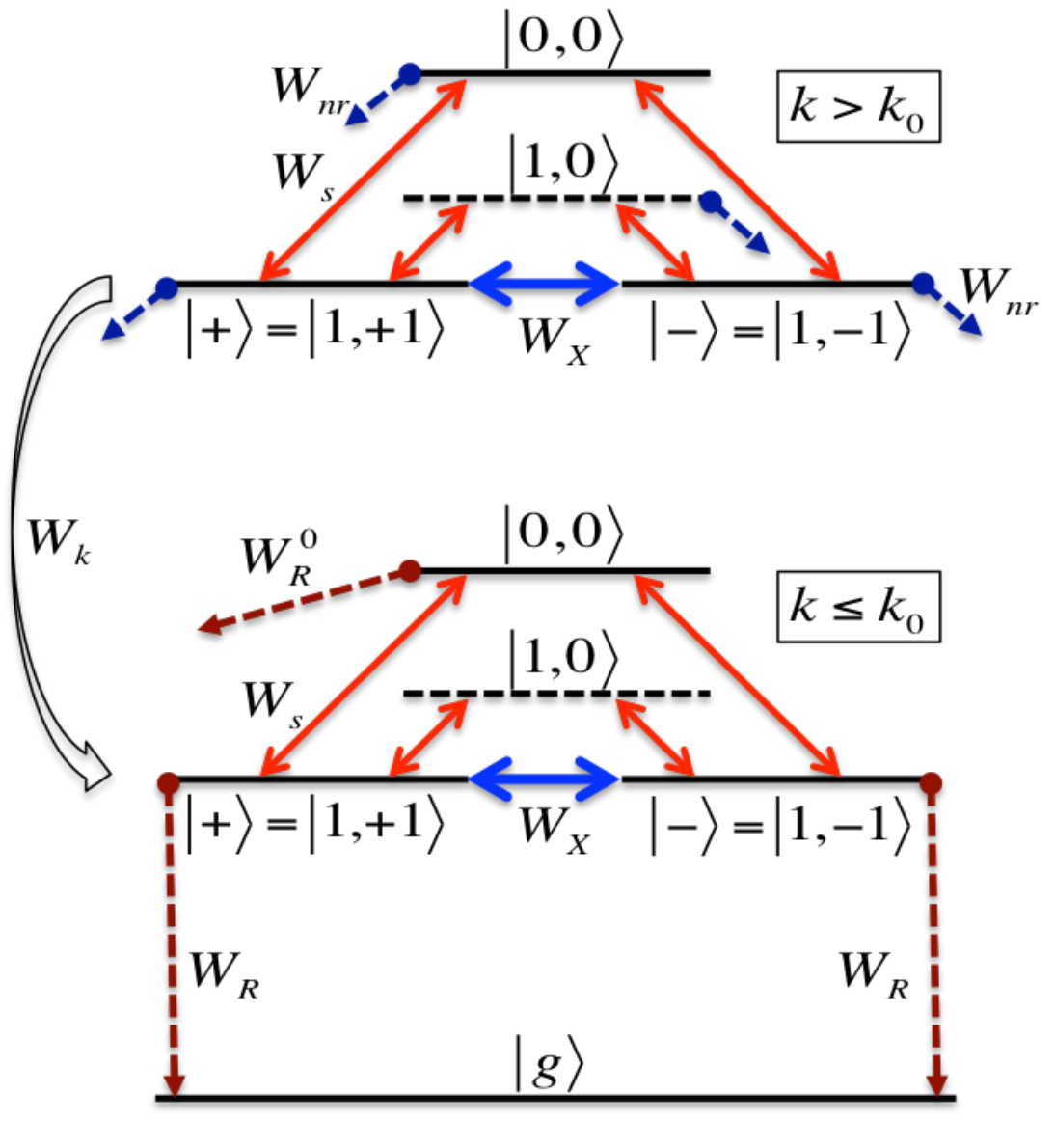}
\caption{Schematic of the model for the exciton dynamics.
}\label{fig:spinmodel} 
\end{figure}

The dynamics of resonantly excited nonthermal excitons in quasi-two-dimensional systems such as GaAs-based quantum-well structures are affected by several physical processes \cite{damen1991,sham1993,maialle1993,vinattieri1994,wang1995,munoz1995,baylac1995,ivchenko1995,amand1997,le-jeune1998,vina1999,amand2008}: (1) momentum relaxation of excitons, (2)  spin relaxation of excitons, and (3) the enhanced radiative recombination and propagation of exciton polaritons. For nonresonantly photoexcited carriers, one should also consider the contribution to dynamics from free carriers, particularly at high temperature. Here, we calculate the population of excitons in various spin and momentum states in GaSe as described in Sec. \ref{sec:bandstructure} using a simplified exciton spin-flip model (Fig. \ref{fig:spinmodel}). The model is adapted from a unified model for resonantly excited excitons in GaAs-based quantum wells \cite{vinattieri1994}. In principle, this model should be valid only for resonant photoexcitation at low temperature where the contributions from free carriers is negligible. Nevertheless, we find that this model reproduces the essential experimental photoluminescence polarization properties and dynamics. 

We label exciton states as $\left|\pm\right>\equiv\left|1,\pm1\right>$, $\left|1, 0\right>$, and $\left|0, 0\right>$. We first consider the case when the excitons are photoexcited in non-radiative, high-momentum states  ($K_\parallel>K_0$) labelled, for example, $\left|+k\right>$, where $K_\parallel$ is the in-plane momentum of excitons and $K_0$ is the photon momentum. Exciton spin-flip (with rate $W_X$) transfers population between $\left|+\right>$ and $\left|-\right>$ states, while electron/hole spin-flip (indistinguishable in our experiments and so characterized by a single rate ($W_s$)) populates the dipole-inactive $\left|1, 0\right>$ non-radiative (dark) state, and the singlet $\left|0, 0\right>$ state (dipole-active for $E\parallel c$). Following Vinattieri et al. \cite{vinattieri1994}, we divide the manifold of $K_\parallel$ states into two sets, one for nearly zero $K_\parallel$ and the other for finite large $K_\parallel$ states. Each set includes the four exciton states. Absorption and emission of acoustic phonons induce transitions between these two sets. We consider only spin-conserving transitions with an effective scattering rate $W_k$. To simplify the model, we neglect any thermal factors associated with the spin-flip rates of excitons and electron/hole.   

The time-dependent population in each state is given by a set of coupled equations:
\begin{equation}
\frac{d}{dt}N_i=M_{ij} \, N_j + G(t) \, \delta_{+1k,i} \,\, ,\\
\end{equation}

\noindent where $N_i$ is the column vector $(N_{+1}, N_{-1}, N_{10}, N_{00}, N_{+1k}, N_{-1k}, N_{10k}, N_{00k})$ and $M$ is a $8\times8$ matrix. $N_{+1}, N_{-1}, N_{10}, N_{00}$ are populations of $K_\parallel \lesssim K_0$ states $\left|+\right>$, $\left|-\right>$, $\left|1, 0\right>$, and $\left|0, 0\right>$, respectively. $N_{+1k}, N_{-1k}, N_{10k}, N_{00k}$ are corresponding $K_\parallel>K_0$ states. $M$ is the following matrix:
\begin{equation}
M=
\begin{bmatrix}
A & C \\
D & B
\end{bmatrix},
\end{equation}
where $A$, $B$, $C$, and $D$ are the following $4\times4$ matrices:

\begin{widetext}
\begin{equation}
A = \left[
\begin{smallmatrix}
-(W_R+W_X+W_s+W_{kp}) & W_X & W_s/4 & W_s/4 \\
W_X & -(W_R+W_X+W_s+W_{kp}) & W_s/4 & W_s/4 \\
W_s/2 & W_s/2 & -(W_s/2+W_{kp}) & 0 \\
W_s/2 & W_s/2 & 0 & -(W_{R}^{0}+W_s/2+W_{kp})
\end{smallmatrix}\right], \nonumber\\
\end{equation}
\begin{equation}
B = \left[
\begin{smallmatrix}
-(W_{nr}+W_X+W_s+W_{km}) & W_X & W_s/4 & W_s/4 \\
W_X & -(W_{nr}+W_X+W_s+W_{km}) & W_s/4 & W_s/4 \\
W_s/2 & W_s/2 & -(W_{nr}+W_s/2+W_{km}) & 0 \\
W_s/2 & W_s/2 & 0 & -(W_{nr}+W_s/2+W_{km})
\end{smallmatrix}\right], \nonumber\\
\end{equation}

\begin{equation}
C=
\begin{bmatrix}
W_{km} & 0 & 0 & 0 \\
0 & W_{km} & 0 & 0 \\
0 & 0 & W_{km} & 0 \\
0 & 0 & 0 & W_{km} \\
\end{bmatrix}, \,
D=
\begin{bmatrix}
W_{kp} & 0 & 0 & 0 \\
0 & W_{kp} & 0 & 0 \\
0 & 0 & W_{kp} & 0 \\
0 & 0 & 0 & W_{kp} \\
\end{bmatrix}. \nonumber
\end{equation}
\end{widetext} 
$W_R$ and $W_R^0$ are the radiative recombination rates for $\left|\pm\right>$ and $\left|0, 0\right>$, respectively. $W_X$ and $W_s$ are the exciton and electron/hole spin-relaxation rates. The acoustic-phonon scattering rates $W_{kp}$ and $W_{km}$ are defined by Vinattieri et al. \cite{vinattieri1994} as 
\begin{align}
W_{kp} &= W_k \, \textrm{exp}\left[-\frac{ \hbar \Gamma_h}{k_B T}\right], \nonumber\\
W_{km} &= W_k \, \left(1-\textrm{exp}\left[-\frac{\hbar \Gamma_h}{k_B T}\right]\right),
\end{align}
where $W_k$ is the effective scattering rate with phonons, and $\hbar \Gamma_h$ is the homogeneous linewidth. The spin-flip rates of electrons and holes cannot be distinguished in this model because spin-flip of electron and holes results in identical transitions within the model.

\begin{figure}[htb]
\centering
\includegraphics[width=0.45 \textwidth]{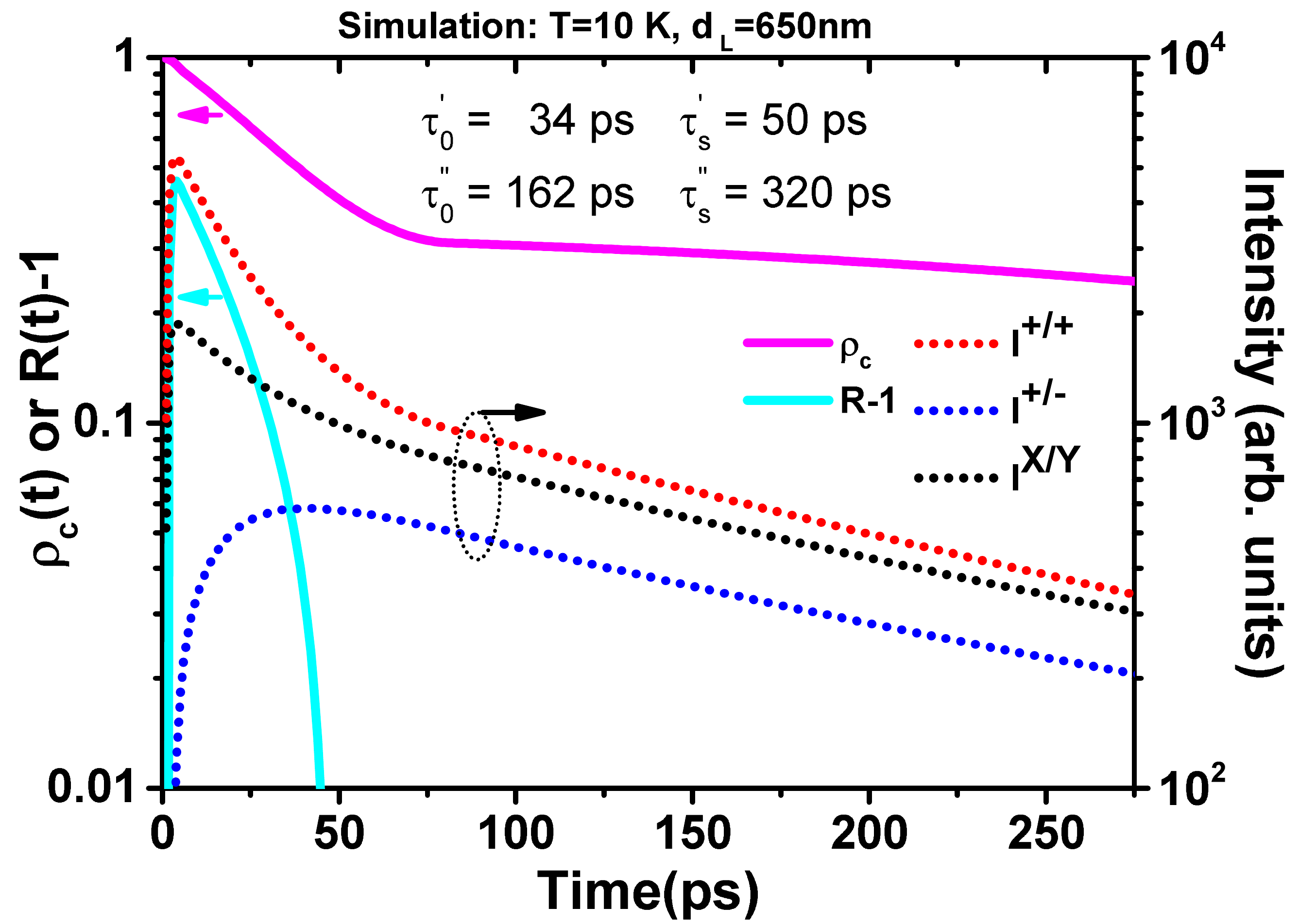}
\caption{Calculated polarized PL dynamics for T = 10 K and $d_L$ = 650 nm.
}\label{fig:t10k_sim} 
\end{figure}

There are six parameters in the model ($W_R^0$, $W_R$, $W_X$, $W_s$, $W_k$, $\Gamma_h$). However, the present and prior experimental data strongly constrain the values these rates can take in fitting the experimental polarized PL dynamics with this model (Fig. \ref{fig:spinmodel}). The radiative recombination rate $W_R^0$ for state $\left|0, 0\right>$ is set to 30 $W_R$ based on the relative absorption coefficients between $\vec{E} \parallel c$ and $\vec{E} \perp c$ light \cite{le-toullec1977}. The value of $\Gamma_h$ is not available but in principle can be measured independently from temperature-dependent PL linewidths \cite{feldmann1987,srinivas1992,citrin1993,aaviksoo1991}. For simplicity, we set $\hbar \Gamma_h = k_B \, T$ because the linewidth increases from $\sim$30 meV at T = 10 K to $\sim$50 meV at T = 300 K. $W_R$ is then largely determined by the decay of population (the measured PL decay). 

We first consider the total time-dependent PL under linearly and circularly polarized excitation conditions. Under linearly polarized excitation, the photoexcited electrons and holes are assumed to be equally distributed over their respective spin states, i.e. $\left|\pm 1/2\right>$, during the initial 2-ps laser excitation. The nongeminate (bimolecular) formation of excitons then produces an initial exciton population distributed equally over the three triplet exciton states $\left|\pm\right>$ and $\left|1, 0\right>$. We contrast this to the case of circularly polarized excitation ($\sigma^+$), under which only the bright exciton state $\left|+\right>$ is initially formed. The total time-dependent populations of $\left|+\right>$ and $\left|-\right>$ created by circularly and linearly polarized light are labeled $I^+(t)$ and $I^X(t)$, respectively. Then the ratio $R(t)=I^+(t)/I^X(t)=\left[I^{+/+}(t)+I^{+/-}(t)\right]/\left[I^{X/X}(t)+I^{X/Y}(t)\right]$ will decrease from approximately 1.5 to 1 as the populations in the three triplet states eventually become nearly equal through spin relaxation under $\sigma^+$ excitation. 

In order to estimate $W_X$, we first focus on polarized PL dynamics under circularly polarized excitation. The sum of time-dependent $\sigma^+$ and $\sigma^-$ PL under circularly polarized $\sigma^\pm$ excitation and $\sigma^X$ and $\sigma^{Y}$ PL under linearly polarized $\sigma^{X/Y}$ excitation are both independent of $W_X$ because $W_X$ does not change the total population in the two radiative states. By fitting experimental results for the polarized and total (i.e., unpolarized) PL versus time, we find that polarized PL dynamics is dominated by $W_s$ at cryogenic temperatures (T = 10 K). 

The preceding discussion of the decay of spin polarization applies to band-edge carriers. However, it does not account for spin relaxation of carriers excited above the gap, which exhibit faster spin relaxation as they cool and scatter into the region $K_\parallel\leq K_0$ \cite{vina1999}. At T = 10 K, the time-dependent degree of circular polarization ($\rho_c(t)$) exhibits biexponential decay. The corresponding time-dependent PL under linearly polarized excitation also decays biexponentially. In addition, the ratio $R(t)$ decays to 1 within about 20 ps. We can account for these experimental features without appeal to the detailed nonequilibrium carrier distributions, by assigning a spin-flip rate that decreases linearly from about $W_s' \approx 500 \ W_s$ to $W_s$ during the thermalization period $\tau_{th}$. The thermalization time can be determined by fitting to the polarized PL dynamics at low temperature ($\tau_{th}\sim$ 80 ps at T = 10 K). After including such phenomenological parameters $W_s' = 500 \ W_s$ and $\tau_{th} = 80$ ps in the model, we reproduce quantitatively $\rho_c(t)$, $I^{+/+}(t)$, $I^{+/-}(t)$, and $I^{X/Y}(t)$, and qualitatively [$R(t)-1$] (Fig.~\ref{fig:t10k_sim}). The parameters used for the calculations are as follows: $W_R$ = 0.006 ps$^{-1}$, $W_{nr} < 10^{-4}$ ps$^{-1}$, $W_X < 10^{-5}$ ps$^{-1}$, $W_s$ = 0.001 ps$^{-1}$, $W_s'$ = 0.5 ps$^{-1}$, $\tau_{th}$ = 80 ps, $\hbar \Gamma_h = 2$ meV, and $W_k$ = 1.0 ps$^{-1}$. The calculated PL dynamics (Fig.~\ref{fig:t10k_sim}) agree with the experimental results (Fig. \ref{fig:ex560-570nm_dyn}). 

This model is also consistent with the remote-edge-luminescence (REL) dynamics reported previously \cite{tang2015a}. Excitation with $\vec{E} \perp c$ light results in linearly polarized emission with polarization along the line from the excitation spot to the emission spot. The highly linearly polarized REL is due to scattering of excitons from the $\left|\pm\right>$ excitons into the $\left|0, 0\right>$ state. The rise of the REL corresponds to the optical spin decay of the PL from the focal spot (with a time constant $\sim$20--30 ps).

\section{Summary}
We have analyzed exciton spin dynamics in GaSe under \emph{nonresonant} circularly polarized optical pumping with a phenomenological exciton spin-flip rate-equation model. The model reproduces the initial unity circular polarization as well as the biexponential decay with a sub-20 ps and a $>$500 ps time constant under nonresonant optical pumping with excess energy up to 0.15 eV at T = 10 K. The separation of the non-degenerate conduction and valence bands from other bands results in angular momentum preservation for both electrons and holes. Spin relaxation of hot spin-polarized carriers nonresonantly optically injected into noncentrosymmetric $\epsilon$-GaSe is dominated by the DP spin-relaxation mechanism; a sub-20 ps spin-relaxation time results. By contrast, the spin-relaxation rate of cold excitons (\emph{e-h} pairs) formed near the $\Gamma$-point near the band edge is greatly reduced as a result of the suppressed DP and EY spin-relaxation mechanisms. 

\begin{acknowledgements}
This work is supported by NSF grant DMR-09055944 and J. Cownen endowment at Michigan State University. This research has used the W. M. Keck Microfabrication Facility. We thank Brage Golding, Bhanu Mahanti, and Carlo Piermarocchi for comments and discussions.
\end{acknowledgements}

\bibliography{/Users/cwlai/Dropbox/Bib/lai_lib}

\begin{thebibliography}{66}%
\makeatletter
\providecommand \@ifxundefined [1]{%
 \@ifx{#1\undefined}
}%
\providecommand \@ifnum [1]{%
 \ifnum #1\expandafter \@firstoftwo
 \else \expandafter \@secondoftwo
 \fi
}%
\providecommand \@ifx [1]{%
 \ifx #1\expandafter \@firstoftwo
 \else \expandafter \@secondoftwo
 \fi
}%
\providecommand \natexlab [1]{#1}%
\providecommand \enquote  [1]{``#1''}%
\providecommand \bibnamefont  [1]{#1}%
\providecommand \bibfnamefont [1]{#1}%
\providecommand \citenamefont [1]{#1}%
\providecommand \href@noop [0]{\@secondoftwo}%
\providecommand \href [0]{\begingroup \@sanitize@url \@href}%
\providecommand \@href[1]{\@@startlink{#1}\@@href}%
\providecommand \@@href[1]{\endgroup#1\@@endlink}%
\providecommand \@sanitize@url [0]{\catcode `\\12\catcode `\$12\catcode
  `\&12\catcode `\#12\catcode `\^12\catcode `\_12\catcode `\%12\relax}%
\providecommand \@@startlink[1]{}%
\providecommand \@@endlink[0]{}%
\providecommand \url  [0]{\begingroup\@sanitize@url \@url }%
\providecommand \@url [1]{\endgroup\@href {#1}{\urlprefix }}%
\providecommand \urlprefix  [0]{URL }%
\providecommand \Eprint [0]{\href }%
\providecommand \doibase [0]{http://dx.doi.org/}%
\providecommand \selectlanguage [0]{\@gobble}%
\providecommand \bibinfo  [0]{\@secondoftwo}%
\providecommand \bibfield  [0]{\@secondoftwo}%
\providecommand \translation [1]{[#1]}%
\providecommand \BibitemOpen [0]{}%
\providecommand \bibitemStop [0]{}%
\providecommand \bibitemNoStop [0]{.\EOS\space}%
\providecommand \EOS [0]{\spacefactor3000\relax}%
\providecommand \BibitemShut  [1]{\csname bibitem#1\endcsname}%
\let\auto@bib@innerbib\@empty
\bibitem [{\citenamefont {D'yakonov}\ and\ \citenamefont
  {Perel'}(1984)}]{dyakonov1984}%
  \BibitemOpen
  \bibfield  {author} {\bibinfo {author} {\bibfnamefont {M.~I.}\ \bibnamefont
  {D'yakonov}}\ and\ \bibinfo {author} {\bibfnamefont {V.~I.}\ \bibnamefont
  {Perel'}},\ }\enquote {\bibinfo {title} {Theory of optical spin orientation
  of electrons and nuclei in semiconductors},}\ in\ \href@noop {} {\emph
  {\bibinfo {booktitle} {Optical Orientation}}},\ \bibinfo {series} {Modern
  Problems in Condensed Matter Sciences}, Vol.~\bibinfo {volume} {8}\ (\bibinfo
   {publisher} {Elsevier},\ \bibinfo {year} {1984})\ pp.\ \bibinfo {pages}
  {11--72}\BibitemShut {NoStop}%
\bibitem [{\citenamefont {Dyakonov}(2008)}]{dyakonov2008}%
  \BibitemOpen
  \bibinfo {editor} {\bibfnamefont {M.~I.}\ \bibnamefont {Dyakonov}},\ ed.,\
  \href {\doibase 10.1007/978-3-540-78820-1} {\emph {\bibinfo {title} {Spin
  Physics in Semiconductors}}},\ \bibinfo {series} {Springer Series in
  Solid-State Science}, Vol.\ \bibinfo {volume} {157}\ (\bibinfo  {publisher}
  {Springer},\ \bibinfo {year} {2008})\BibitemShut {NoStop}%
\bibitem [{\citenamefont {Pikus}\ and\ \citenamefont
  {Titkov}(1984)}]{pikus1984}%
  \BibitemOpen
  \bibfield  {author} {\bibinfo {author} {\bibfnamefont {G.~E.}\ \bibnamefont
  {Pikus}}\ and\ \bibinfo {author} {\bibfnamefont {A.~N.}\ \bibnamefont
  {Titkov}},\ }\enquote {\bibinfo {title} {Spin relaxation under optical
  orientation in semiconductors},}\ in\ \href@noop {} {\emph {\bibinfo
  {booktitle} {Optical Orientation}}},\ Vol.~\bibinfo {volume} {8}\ (\bibinfo
  {publisher} {Elsevier},\ \bibinfo {year} {1984})\ pp.\ \bibinfo {pages}
  {73--131}\BibitemShut {NoStop}%
\bibitem [{\citenamefont {Wu}\ \emph {et~al.}(2010)\citenamefont {Wu},
  \citenamefont {Jiang},\ and\ \citenamefont {Weng}}]{wu2010}%
  \BibitemOpen
  \bibfield  {author} {\bibinfo {author} {\bibfnamefont {M.~W.}\ \bibnamefont
  {Wu}}, \bibinfo {author} {\bibfnamefont {J.~H.}\ \bibnamefont {Jiang}}, \
  and\ \bibinfo {author} {\bibfnamefont {M.~Q.}\ \bibnamefont {Weng}},\
  }\bibfield  {title} {\enquote {\bibinfo {title} {Spin dynamics in
  semiconductors},}\ }\href {\doibase 10.1016/j.physrep.2010.04.002} {\bibfield
   {journal} {\bibinfo  {journal} {Phys. Rep.}\ }\textbf {\bibinfo {volume}
  {493}},\ \bibinfo {pages} {61 -- 236} (\bibinfo {year} {2010})}\BibitemShut
  {NoStop}%
\bibitem [{\citenamefont {{\v Z}uti{\'c}}\ \emph {et~al.}(2004)\citenamefont
  {{\v Z}uti{\'c}}, \citenamefont {Fabian},\ and\ \citenamefont
  {Das~Sarma}}]{zutic2004}%
  \BibitemOpen
  \bibfield  {author} {\bibinfo {author} {\bibfnamefont {I.}~\bibnamefont {{\v
  Z}uti{\'c}}}, \bibinfo {author} {\bibfnamefont {J.}~\bibnamefont {Fabian}}, \
  and\ \bibinfo {author} {\bibfnamefont {S.}~\bibnamefont {Das~Sarma}},\
  }\bibfield  {title} {\enquote {\bibinfo {title} {Spintronics: Fundamentals
  and applications},}\ }\href {\doibase 10.1103/RevModPhys.76.323} {\bibfield
  {journal} {\bibinfo  {journal} {Rev. Mod. Phys.}\ }\textbf {\bibinfo {volume}
  {76}},\ \bibinfo {pages} {323--410} (\bibinfo {year} {2004})}\BibitemShut
  {NoStop}%
\bibitem [{\citenamefont {Awschalom}\ \emph {et~al.}(2013)\citenamefont
  {Awschalom}, \citenamefont {Bassett}, \citenamefont {Dzurak}, \citenamefont
  {Hu},\ and\ \citenamefont {Petta}}]{awschalom2013}%
  \BibitemOpen
  \bibfield  {author} {\bibinfo {author} {\bibfnamefont {D.~D.}\ \bibnamefont
  {Awschalom}}, \bibinfo {author} {\bibfnamefont {L.~C.}\ \bibnamefont
  {Bassett}}, \bibinfo {author} {\bibfnamefont {A.~S.}\ \bibnamefont {Dzurak}},
  \bibinfo {author} {\bibfnamefont {E.~L.}\ \bibnamefont {Hu}}, \ and\ \bibinfo
  {author} {\bibfnamefont {J.~R.}\ \bibnamefont {Petta}},\ }\bibfield  {title}
  {\enquote {\bibinfo {title} {Quantum spintronics: engineering and
  manipulating atom-like spins in semiconductors.}}\ }\href {\doibase
  10.1126/science.1231364} {\bibfield  {journal} {\bibinfo  {journal}
  {Science}\ }\textbf {\bibinfo {volume} {339}},\ \bibinfo {pages} {1174--9}
  (\bibinfo {year} {2013})}\BibitemShut {NoStop}%
\bibitem [{\citenamefont {Kikkawa}\ \emph {et~al.}(1997)\citenamefont
  {Kikkawa}, \citenamefont {Smorchkova}, \citenamefont {Samarth},\ and\
  \citenamefont {Awschalom}}]{kikkawa1997}%
  \BibitemOpen
  \bibfield  {author} {\bibinfo {author} {\bibfnamefont {J.~M.}\ \bibnamefont
  {Kikkawa}}, \bibinfo {author} {\bibfnamefont {I.~P.}\ \bibnamefont
  {Smorchkova}}, \bibinfo {author} {\bibfnamefont {N.}~\bibnamefont {Samarth}},
  \ and\ \bibinfo {author} {\bibfnamefont {D.~D.}\ \bibnamefont {Awschalom}},\
  }\bibfield  {title} {\enquote {\bibinfo {title} {Room-temperature spin memory
  in two-dimensional electron gases},}\ }\href {\doibase
  10.1126/science.277.5330.1284} {\bibfield  {journal} {\bibinfo  {journal}
  {Science}\ }\textbf {\bibinfo {volume} {277}},\ \bibinfo {pages} {1284--1287}
  (\bibinfo {year} {1997})}\BibitemShut {NoStop}%
\bibitem [{\citenamefont {Ohno}\ \emph {et~al.}(1999)\citenamefont {Ohno},
  \citenamefont {Terauchi}, \citenamefont {Adachi}, \citenamefont {Matsukura},\
  and\ \citenamefont {Ohno}}]{ohno1999}%
  \BibitemOpen
  \bibfield  {author} {\bibinfo {author} {\bibfnamefont {Y.}~\bibnamefont
  {Ohno}}, \bibinfo {author} {\bibfnamefont {R.}~\bibnamefont {Terauchi}},
  \bibinfo {author} {\bibfnamefont {T.}~\bibnamefont {Adachi}}, \bibinfo
  {author} {\bibfnamefont {F.}~\bibnamefont {Matsukura}}, \ and\ \bibinfo
  {author} {\bibfnamefont {H.}~\bibnamefont {Ohno}},\ }\bibfield  {title}
  {\enquote {\bibinfo {title} {Spin relaxation in {GaAs(110)} quantum wells},}\
  }\href {\doibase 10.1103/PhysRevLett.83.4196} {\bibfield  {journal} {\bibinfo
   {journal} {Phys. Rev. Lett.}\ }\textbf {\bibinfo {volume} {83}},\ \bibinfo
  {pages} {4196--4199} (\bibinfo {year} {1999})}\BibitemShut {NoStop}%
\bibitem [{\citenamefont {Kohl}\ \emph {et~al.}(1991)\citenamefont {Kohl},
  \citenamefont {Freeman}, \citenamefont {Awschalom},\ and\ \citenamefont
  {Hong}}]{kohl1991}%
  \BibitemOpen
  \bibfield  {author} {\bibinfo {author} {\bibfnamefont {M.}~\bibnamefont
  {Kohl}}, \bibinfo {author} {\bibfnamefont {M.~R.}\ \bibnamefont {Freeman}},
  \bibinfo {author} {\bibfnamefont {D.~D.}\ \bibnamefont {Awschalom}}, \ and\
  \bibinfo {author} {\bibfnamefont {J.~M.}\ \bibnamefont {Hong}},\ }\bibfield
  {title} {\enquote {\bibinfo {title} {Femtosecond spectroscopy of carrier-spin
  relaxation in {GaAs-Al$_x$Ga$_{1-x}$As} quantum wells},}\ }\href {\doibase
  10.1103/PhysRevB.44.5923} {\bibfield  {journal} {\bibinfo  {journal} {Phys.
  Rev. B}\ }\textbf {\bibinfo {volume} {44}},\ \bibinfo {pages} {5923}
  (\bibinfo {year} {1991})}\BibitemShut {NoStop}%
\bibitem [{\citenamefont {Amand}\ \emph {et~al.}(1994)\citenamefont {Amand},
  \citenamefont {Dareys}, \citenamefont {Baylac}, \citenamefont {Marie},
  \citenamefont {Barrau}, \citenamefont {Brousseau}, \citenamefont {Dunstan},\
  and\ \citenamefont {Planel}}]{amand1994}%
  \BibitemOpen
  \bibfield  {author} {\bibinfo {author} {\bibfnamefont {T.}~\bibnamefont
  {Amand}}, \bibinfo {author} {\bibfnamefont {B.}~\bibnamefont {Dareys}},
  \bibinfo {author} {\bibfnamefont {B.}~\bibnamefont {Baylac}}, \bibinfo
  {author} {\bibfnamefont {X.}~\bibnamefont {Marie}}, \bibinfo {author}
  {\bibfnamefont {J.}~\bibnamefont {Barrau}}, \bibinfo {author} {\bibfnamefont
  {M.}~\bibnamefont {Brousseau}}, \bibinfo {author} {\bibfnamefont {D.~J.}\
  \bibnamefont {Dunstan}}, \ and\ \bibinfo {author} {\bibfnamefont
  {R.}~\bibnamefont {Planel}},\ }\bibfield  {title} {\enquote {\bibinfo {title}
  {Exciton formation and hole-spin relaxation in intrinsic quantum wells},}\
  }\href {\doibase 10.1103/PhysRevB.50.11624} {\bibfield  {journal} {\bibinfo
  {journal} {Phys. Rev. B}\ }\textbf {\bibinfo {volume} {50}},\ \bibinfo
  {pages} {11624--11628} (\bibinfo {year} {1994})}\BibitemShut {NoStop}%
\bibitem [{\citenamefont {Pfalz}\ \emph {et~al.}(2005)\citenamefont {Pfalz},
  \citenamefont {Winkler}, \citenamefont {Nowitzki}, \citenamefont {Reuter},
  \citenamefont {Wieck}, \citenamefont {H{\"a}gele},\ and\ \citenamefont
  {Oestreich}}]{pfalz2005}%
  \BibitemOpen
  \bibfield  {author} {\bibinfo {author} {\bibfnamefont {S.}~\bibnamefont
  {Pfalz}}, \bibinfo {author} {\bibfnamefont {R.}~\bibnamefont {Winkler}},
  \bibinfo {author} {\bibfnamefont {T.}~\bibnamefont {Nowitzki}}, \bibinfo
  {author} {\bibfnamefont {D.}~\bibnamefont {Reuter}}, \bibinfo {author}
  {\bibfnamefont {A.~D.}\ \bibnamefont {Wieck}}, \bibinfo {author}
  {\bibfnamefont {D.}~\bibnamefont {H{\"a}gele}}, \ and\ \bibinfo {author}
  {\bibfnamefont {M.}~\bibnamefont {Oestreich}},\ }\bibfield  {title} {\enquote
  {\bibinfo {title} {Optical orientation of electron spins in {GaAs} quantum
  wells},}\ }\href {\doibase 10.1103/PhysRevB.71.165305} {\bibfield  {journal}
  {\bibinfo  {journal} {Phys. Rev. B}\ }\textbf {\bibinfo {volume} {71}},\
  \bibinfo {pages} {165305} (\bibinfo {year} {2005})}\BibitemShut {NoStop}%
\bibitem [{\citenamefont {Tang}\ \emph
  {et~al.}(2015{\natexlab{a}})\citenamefont {Tang}, \citenamefont {Xie},
  \citenamefont {Mandal}, \citenamefont {McGuire},\ and\ \citenamefont
  {Lai}}]{tang2015}%
  \BibitemOpen
  \bibfield  {author} {\bibinfo {author} {\bibfnamefont {Y.}~\bibnamefont
  {Tang}}, \bibinfo {author} {\bibfnamefont {W.}~\bibnamefont {Xie}}, \bibinfo
  {author} {\bibfnamefont {K.~C.}\ \bibnamefont {Mandal}}, \bibinfo {author}
  {\bibfnamefont {J.~A.}\ \bibnamefont {McGuire}}, \ and\ \bibinfo {author}
  {\bibfnamefont {C.~W.}\ \bibnamefont {Lai}},\ }\bibfield  {title} {\enquote
  {\bibinfo {title} {Optical and spin polarization dynamics in {GaSe}
  nanoslabs},}\ }\href {\doibase 10.1103/physrevb.91.195429} {\bibfield
  {journal} {\bibinfo  {journal} {Phys. Rev. B}\ }\textbf {\bibinfo {volume}
  {91}},\ \bibinfo {pages} {195429} (\bibinfo {year}
  {2015}{\natexlab{a}})}\BibitemShut {NoStop}%
\bibitem [{\citenamefont {Gamarts}\ \emph {et~al.}(1977)\citenamefont
  {Gamarts}, \citenamefont {Ivchenko}, \citenamefont {Karaman}, \citenamefont
  {Mushinskii}, \citenamefont {Pikus}, \citenamefont {Razbirin},\ and\
  \citenamefont {Starukhin}}]{gamarts1977}%
  \BibitemOpen
  \bibfield  {author} {\bibinfo {author} {\bibfnamefont {E.~M.}\ \bibnamefont
  {Gamarts}}, \bibinfo {author} {\bibfnamefont {E.~L.}\ \bibnamefont
  {Ivchenko}}, \bibinfo {author} {\bibfnamefont {M.~I.}\ \bibnamefont
  {Karaman}}, \bibinfo {author} {\bibfnamefont {V.~P.}\ \bibnamefont
  {Mushinskii}}, \bibinfo {author} {\bibfnamefont {G.~E.}\ \bibnamefont
  {Pikus}}, \bibinfo {author} {\bibfnamefont {B.~S.}\ \bibnamefont {Razbirin}},
  \ and\ \bibinfo {author} {\bibfnamefont {A.~N.}\ \bibnamefont {Starukhin}},\
  }\bibfield  {title} {\enquote {\bibinfo {title} {Optical orientation and
  alignment of free excitons in {GaSe} during resonance excitation.
  experiment},}\ }\href@noop {} {\bibfield  {journal} {\bibinfo  {journal}
  {Sov. Phys. JETP}\ }\textbf {\bibinfo {volume} {46}},\ \bibinfo {pages} {590}
  (\bibinfo {year} {1977})}\BibitemShut {NoStop}%
\bibitem [{\citenamefont {Ivchenko}\ \emph {et~al.}(1977)\citenamefont
  {Ivchenko}, \citenamefont {Pikus}, \citenamefont {Razbirin},\ and\
  \citenamefont {Starukhin}}]{ivchenko1977}%
  \BibitemOpen
  \bibfield  {author} {\bibinfo {author} {\bibfnamefont {E.~L.}\ \bibnamefont
  {Ivchenko}}, \bibinfo {author} {\bibfnamefont {G.~E.}\ \bibnamefont {Pikus}},
  \bibinfo {author} {\bibfnamefont {B.~S.}\ \bibnamefont {Razbirin}}, \ and\
  \bibinfo {author} {\bibfnamefont {A.~I.}\ \bibnamefont {Starukhin}},\
  }\bibfield  {title} {\enquote {\bibinfo {title} {Optical orientation and
  alignment of free excitons in {GaSe} under resonant excitation. theory.}}\
  }\href@noop {} {\bibfield  {journal} {\bibinfo  {journal} {Sov. Phys. JETP}\
  }\textbf {\bibinfo {volume} {45}},\ \bibinfo {pages} {1172--1180} (\bibinfo
  {year} {1977})}\BibitemShut {NoStop}%
\bibitem [{\citenamefont {Do}\ \emph {et~al.}(2015)\citenamefont {Do},
  \citenamefont {Mahanti},\ and\ \citenamefont {Lai}}]{do2015}%
  \BibitemOpen
  \bibfield  {author} {\bibinfo {author} {\bibfnamefont {D.~T.}\ \bibnamefont
  {Do}}, \bibinfo {author} {\bibfnamefont {S.~D.}\ \bibnamefont {Mahanti}}, \
  and\ \bibinfo {author} {\bibfnamefont {C.~W.}\ \bibnamefont {Lai}},\
  }\bibfield  {title} {\enquote {\bibinfo {title} {Spin splitting in {2D}
  monochalcogenide semiconductors},}\ }\href {http://arxiv.org/abs/1504.00725}
  {\ ,\ \bibinfo {pages} {arXiv:1504.00725} (\bibinfo {year}
  {2015})}\BibitemShut {NoStop}%
\bibitem [{\citenamefont {Brebner}\ and\ \citenamefont
  {Mooser}(1967)}]{brebner1967}%
  \BibitemOpen
  \bibfield  {author} {\bibinfo {author} {\bibfnamefont {J.~L.}\ \bibnamefont
  {Brebner}}\ and\ \bibinfo {author} {\bibfnamefont {E.}~\bibnamefont
  {Mooser}},\ }\bibfield  {title} {\enquote {\bibinfo {title} {Excitons in
  {GaSe} polytypes},}\ }\href {\doibase 10.1016/0375-9601(67)90433-1}
  {\bibfield  {journal} {\bibinfo  {journal} {Phys. Lett. A}\ }\textbf
  {\bibinfo {volume} {24}},\ \bibinfo {pages} {274--275} (\bibinfo {year}
  {1967})}\BibitemShut {NoStop}%
\bibitem [{\citenamefont {Aulich}\ \emph {et~al.}(1969)\citenamefont {Aulich},
  \citenamefont {Brebner},\ and\ \citenamefont {Mooser}}]{aulich1969}%
  \BibitemOpen
  \bibfield  {author} {\bibinfo {author} {\bibfnamefont {E.}~\bibnamefont
  {Aulich}}, \bibinfo {author} {\bibfnamefont {J.~L. .~L.}\ \bibnamefont
  {Brebner}}, \ and\ \bibinfo {author} {\bibfnamefont {E.}~\bibnamefont
  {Mooser}},\ }\bibfield  {title} {\enquote {\bibinfo {title} {Indirect energy
  gap in {GaSe} and {GaS}},}\ }\href {\doibase 10.1002/pssb.19690310115}
  {\bibfield  {journal} {\bibinfo  {journal} {Phys. Status Solidi B}\ }\textbf
  {\bibinfo {volume} {31}},\ \bibinfo {pages} {129--131} (\bibinfo {year}
  {1969})}\BibitemShut {NoStop}%
\bibitem [{\citenamefont {Mercier}\ \emph {et~al.}(1973)\citenamefont
  {Mercier}, \citenamefont {Mooser},\ and\ \citenamefont
  {Voitchovsky}}]{mercier1973}%
  \BibitemOpen
  \bibfield  {author} {\bibinfo {author} {\bibfnamefont {A.}~\bibnamefont
  {Mercier}}, \bibinfo {author} {\bibfnamefont {E.}~\bibnamefont {Mooser}}, \
  and\ \bibinfo {author} {\bibfnamefont {J.~P.}\ \bibnamefont {Voitchovsky}},\
  }\bibfield  {title} {\enquote {\bibinfo {title} {Near edge optical absorption
  and luminescence of {GaSe}, {GaS} and of mixed crystals},}\ }\href {\doibase
  10.1016/0022-2313(73)90070-7} {\bibfield  {journal} {\bibinfo  {journal} {J.
  Lumin.}\ }\textbf {\bibinfo {volume} {7}},\ \bibinfo {pages} {241 -- 266}
  (\bibinfo {year} {1973})}\BibitemShut {NoStop}%
\bibitem [{\citenamefont {Mooser}\ and\ \citenamefont
  {Schl{\"u}ter}(1973)}]{mooser1973}%
  \BibitemOpen
  \bibfield  {author} {\bibinfo {author} {\bibfnamefont {E.}~\bibnamefont
  {Mooser}}\ and\ \bibinfo {author} {\bibfnamefont {M.}~\bibnamefont
  {Schl{\"u}ter}},\ }\bibfield  {title} {\enquote {\bibinfo {title} {The
  band-gap excitons in gallium selenide},}\ }\href {\doibase
  10.1007/BF02832647} {\bibfield  {journal} {\bibinfo  {journal} {Nuovo Cimento
  B}\ }\textbf {\bibinfo {volume} {18}},\ \bibinfo {pages} {164--208} (\bibinfo
  {year} {1973})}\BibitemShut {NoStop}%
\bibitem [{\citenamefont {Schl\"{u}ter}\ \emph {et~al.}(1976)\citenamefont
  {Schl\"{u}ter}, \citenamefont {Camassel}, \citenamefont {Kohn}, \citenamefont
  {Voitchovsky}, \citenamefont {Shen},\ and\ \citenamefont
  {Cohen}}]{schluter1976}%
  \BibitemOpen
  \bibfield  {author} {\bibinfo {author} {\bibfnamefont {M.}~\bibnamefont
  {Schl\"{u}ter}}, \bibinfo {author} {\bibfnamefont {J.}~\bibnamefont
  {Camassel}}, \bibinfo {author} {\bibfnamefont {S.}~\bibnamefont {Kohn}},
  \bibinfo {author} {\bibfnamefont {J.~P.}\ \bibnamefont {Voitchovsky}},
  \bibinfo {author} {\bibfnamefont {Y.~R.}\ \bibnamefont {Shen}}, \ and\
  \bibinfo {author} {\bibfnamefont {Marvin~L.}\ \bibnamefont {Cohen}},\
  }\bibfield  {title} {\enquote {\bibinfo {title} {Optical properties of {GaSe}
  and {GaS$_x$Se$_{1-x}$} mixed crystals},}\ }\href {\doibase
  10.1103/PhysRevB.13.3534} {\bibfield  {journal} {\bibinfo  {journal} {Phys.
  Rev. B}\ }\textbf {\bibinfo {volume} {13}},\ \bibinfo {pages} {3534--3547}
  (\bibinfo {year} {1976})}\BibitemShut {NoStop}%
\bibitem [{\citenamefont {Le~Toullec}\ \emph {et~al.}(1977)\citenamefont
  {Le~Toullec}, \citenamefont {Piccioli}, \citenamefont {Mejatty},\ and\
  \citenamefont {Balkanski}}]{le-toullec1977}%
  \BibitemOpen
  \bibfield  {author} {\bibinfo {author} {\bibfnamefont {R.}~\bibnamefont
  {Le~Toullec}}, \bibinfo {author} {\bibfnamefont {N.}~\bibnamefont
  {Piccioli}}, \bibinfo {author} {\bibfnamefont {M.}~\bibnamefont {Mejatty}}, \
  and\ \bibinfo {author} {\bibfnamefont {M.}~\bibnamefont {Balkanski}},\
  }\bibfield  {title} {\enquote {\bibinfo {title} {Optical constants of
  {$\epsilon$-GaSe}},}\ }\href {\doibase 10.1007/BF02723483} {\bibfield
  {journal} {\bibinfo  {journal} {Nuovo Cimento B}\ }\textbf {\bibinfo {volume}
  {38}},\ \bibinfo {pages} {159--167} (\bibinfo {year} {1977})}\BibitemShut
  {NoStop}%
\bibitem [{\citenamefont {Le~Toullec}\ \emph {et~al.}(1980)\citenamefont
  {Le~Toullec}, \citenamefont {Piccioli},\ and\ \citenamefont
  {Chervin}}]{le-toullec1980}%
  \BibitemOpen
  \bibfield  {author} {\bibinfo {author} {\bibfnamefont {R.}~\bibnamefont
  {Le~Toullec}}, \bibinfo {author} {\bibfnamefont {N.}~\bibnamefont
  {Piccioli}}, \ and\ \bibinfo {author} {\bibfnamefont {J.~C.}\ \bibnamefont
  {Chervin}},\ }\bibfield  {title} {\enquote {\bibinfo {title} {Optical
  properties of the band-edge exciton in {GaSe} crystals at 10 {K}},}\ }\href
  {\doibase 10.1103/PhysRevB.22.6162} {\bibfield  {journal} {\bibinfo
  {journal} {Phys. Rev. B}\ }\textbf {\bibinfo {volume} {22}},\ \bibinfo
  {pages} {6162} (\bibinfo {year} {1980})}\BibitemShut {NoStop}%
\bibitem [{\citenamefont {Sasaki}\ and\ \citenamefont
  {Nishina}(1981)}]{sasaki1981}%
  \BibitemOpen
  \bibfield  {author} {\bibinfo {author} {\bibfnamefont {Y.}~\bibnamefont
  {Sasaki}}\ and\ \bibinfo {author} {\bibfnamefont {Y.}~\bibnamefont
  {Nishina}},\ }\bibfield  {title} {\enquote {\bibinfo {title}
  {Photoluminescence studies of indirect bound excitons in
  {$\epsilon$-GaSe}},}\ }\href {\doibase 10.1103/PhysRevB.23.4089} {\bibfield
  {journal} {\bibinfo  {journal} {Phys. Rev. B}\ }\textbf {\bibinfo {volume}
  {23}},\ \bibinfo {pages} {4089--4096} (\bibinfo {year} {1981})}\BibitemShut
  {NoStop}%
\bibitem [{\citenamefont {Capozzi}\ \emph {et~al.}(1993)\citenamefont
  {Capozzi}, \citenamefont {Pavesi},\ and\ \citenamefont
  {Staehli}}]{capozzi1993}%
  \BibitemOpen
  \bibfield  {author} {\bibinfo {author} {\bibfnamefont {V.}~\bibnamefont
  {Capozzi}}, \bibinfo {author} {\bibfnamefont {L.}~\bibnamefont {Pavesi}}, \
  and\ \bibinfo {author} {\bibfnamefont {J.~L.}\ \bibnamefont {Staehli}},\
  }\bibfield  {title} {\enquote {\bibinfo {title} {Exciton-carrier scattering
  in gallium selenide},}\ }\href {\doibase 10.1103/PhysRevB.47.6340} {\bibfield
   {journal} {\bibinfo  {journal} {Phys. Rev. B.}\ }\textbf {\bibinfo {volume}
  {47}},\ \bibinfo {pages} {6340--6349} (\bibinfo {year} {1993})}\BibitemShut
  {NoStop}%
\bibitem [{\citenamefont {Sasaki}\ \emph
  {et~al.}(1975{\natexlab{a}})\citenamefont {Sasaki}, \citenamefont
  {Hamaguchi},\ and\ \citenamefont {Nakai}}]{sasaki1975a}%
  \BibitemOpen
  \bibfield  {author} {\bibinfo {author} {\bibfnamefont {Y.}~\bibnamefont
  {Sasaki}}, \bibinfo {author} {\bibfnamefont {C.}~\bibnamefont {Hamaguchi}}, \
  and\ \bibinfo {author} {\bibfnamefont {J.}~\bibnamefont {Nakai}},\ }\bibfield
   {title} {\enquote {\bibinfo {title} {Electroreflectance of {GaSe. I. A}round
  3.4 {eV}},}\ }\href {\doibase 10.1143/JPSJ.38.162} {\bibfield  {journal}
  {\bibinfo  {journal} {J. Phys. Soc. Jpn.}\ }\textbf {\bibinfo {volume}
  {38}},\ \bibinfo {pages} {162--168} (\bibinfo {year}
  {1975}{\natexlab{a}})}\BibitemShut {NoStop}%
\bibitem [{\citenamefont {Sasaki}\ \emph
  {et~al.}(1975{\natexlab{b}})\citenamefont {Sasaki}, \citenamefont
  {Hamaguchi},\ and\ \citenamefont {Nakai}}]{sasaki1975b}%
  \BibitemOpen
  \bibfield  {author} {\bibinfo {author} {\bibfnamefont {Y.}~\bibnamefont
  {Sasaki}}, \bibinfo {author} {\bibfnamefont {C.}~\bibnamefont {Hamaguchi}}, \
  and\ \bibinfo {author} {\bibfnamefont {J.}~\bibnamefont {Nakai}},\ }\bibfield
   {title} {\enquote {\bibinfo {title} {Electroreflectance of {GaSe. II.}
  3.5-4.1 e{V} region},}\ }\href {\doibase 10.1143/JPSJ.38.169} {\bibfield
  {journal} {\bibinfo  {journal} {J. Phys. Soc. Jpn.}\ }\textbf {\bibinfo
  {volume} {38}},\ \bibinfo {pages} {169--174} (\bibinfo {year}
  {1975}{\natexlab{b}})}\BibitemShut {NoStop}%
\bibitem [{\citenamefont {Mandal}\ \emph {et~al.}(2008)\citenamefont {Mandal},
  \citenamefont {Mertiri}, \citenamefont {Pabst}, \citenamefont {Roy},
  \citenamefont {Cui}, \citenamefont {Battacharya}, \citenamefont {Groza},
  \citenamefont {Burger}, \citenamefont {Conway},\ and\ \citenamefont
  {Nikolic}}]{mandal2008a}%
  \BibitemOpen
  \bibfield  {author} {\bibinfo {author} {\bibfnamefont {K.~C.}\ \bibnamefont
  {Mandal}}, \bibinfo {author} {\bibfnamefont {A.}~\bibnamefont {Mertiri}},
  \bibinfo {author} {\bibfnamefont {G.~W.}\ \bibnamefont {Pabst}}, \bibinfo
  {author} {\bibfnamefont {R.~G.}\ \bibnamefont {Roy}}, \bibinfo {author}
  {\bibfnamefont {Y.}~\bibnamefont {Cui}}, \bibinfo {author} {\bibfnamefont
  {P.}~\bibnamefont {Battacharya}}, \bibinfo {author} {\bibfnamefont
  {M.}~\bibnamefont {Groza}}, \bibinfo {author} {\bibfnamefont
  {A.}~\bibnamefont {Burger}}, \bibinfo {author} {\bibfnamefont {A.~M.}\
  \bibnamefont {Conway}}, \ and\ \bibinfo {author} {\bibfnamefont {R.~J.}\
  \bibnamefont {Nikolic}},\ }\bibfield  {title} {\enquote {\bibinfo {title}
  {Layered {III-VI} chalcogenide semiconductor crystals for radiation
  detectors},}\ }in\ \href {\doibase 10.1117/12.796235} {\emph {\bibinfo
  {booktitle} {Proc. SPIE}}},\ Vol.\ \bibinfo {volume} {7079}\ (\bibinfo {year}
  {2008})\ p.\ \bibinfo {pages} {70790O}\BibitemShut {NoStop}%
\bibitem [{\citenamefont {Pavesi}\ \emph {et~al.}(1989)\citenamefont {Pavesi},
  \citenamefont {Staehli},\ and\ \citenamefont {Capozzi}}]{pavesi1989}%
  \BibitemOpen
  \bibfield  {author} {\bibinfo {author} {\bibfnamefont {L.}~\bibnamefont
  {Pavesi}}, \bibinfo {author} {\bibfnamefont {J.~L.}\ \bibnamefont {Staehli}},
  \ and\ \bibinfo {author} {\bibfnamefont {V.}~\bibnamefont {Capozzi}},\
  }\bibfield  {title} {\enquote {\bibinfo {title} {Mott transition of the
  excitons in {GaSe}},}\ }\href {\doibase 10.1103/PhysRevB.39.10982} {\bibfield
   {journal} {\bibinfo  {journal} {Phys. Rev. B}\ }\textbf {\bibinfo {volume}
  {39}},\ \bibinfo {pages} {10982--10994} (\bibinfo {year} {1989})}\BibitemShut
  {NoStop}%
\bibitem [{\citenamefont {Piccioli}\ \emph {et~al.}(1977)\citenamefont
  {Piccioli}, \citenamefont {Le~Toullec}, \citenamefont {Mejatty},\ and\
  \citenamefont {Balkanski}}]{piccioli1977}%
  \BibitemOpen
  \bibfield  {author} {\bibinfo {author} {\bibfnamefont {N.}~\bibnamefont
  {Piccioli}}, \bibinfo {author} {\bibfnamefont {R.}~\bibnamefont
  {Le~Toullec}}, \bibinfo {author} {\bibfnamefont {M.}~\bibnamefont {Mejatty}},
  \ and\ \bibinfo {author} {\bibfnamefont {M.}~\bibnamefont {Balkanski}},\
  }\bibfield  {title} {\enquote {\bibinfo {title} {Refractive index of {GaSe}
  between 0.45 $\mu$m and 330 $\mu$m},}\ }\href {\doibase 10.1364/AO.16.001236}
  {\bibfield  {journal} {\bibinfo  {journal} {Appl. Opt.}\ }\textbf {\bibinfo
  {volume} {16}},\ \bibinfo {pages} {1236--1238} (\bibinfo {year}
  {1977})}\BibitemShut {NoStop}%
\bibitem [{\citenamefont {Adachi}\ and\ \citenamefont
  {Shindo}(1992)}]{adachi1992}%
  \BibitemOpen
  \bibfield  {author} {\bibinfo {author} {\bibfnamefont {S.}~\bibnamefont
  {Adachi}}\ and\ \bibinfo {author} {\bibfnamefont {Y.}~\bibnamefont
  {Shindo}},\ }\bibfield  {title} {\enquote {\bibinfo {title} {Optical
  constants of {$\epsilon$-GaSe}},}\ }\href {\doibase 10.1063/1.351362}
  {\bibfield  {journal} {\bibinfo  {journal} {J. Appl. Phys.}\ }\textbf
  {\bibinfo {volume} {71}},\ \bibinfo {pages} {428--431} (\bibinfo {year}
  {1992})}\BibitemShut {NoStop}%
\bibitem [{\citenamefont {Tang}\ \emph
  {et~al.}(2015{\natexlab{b}})\citenamefont {Tang}, \citenamefont {Xie},
  \citenamefont {Mandal}, \citenamefont {McGuire},\ and\ \citenamefont
  {Lai}}]{tang2015a}%
  \BibitemOpen
  \bibfield  {author} {\bibinfo {author} {\bibfnamefont {Y.}~\bibnamefont
  {Tang}}, \bibinfo {author} {\bibfnamefont {W.}~\bibnamefont {Xie}}, \bibinfo
  {author} {\bibfnamefont {K.~C.}\ \bibnamefont {Mandal}}, \bibinfo {author}
  {\bibfnamefont {J.~A.}\ \bibnamefont {McGuire}}, \ and\ \bibinfo {author}
  {\bibfnamefont {C.~W.}\ \bibnamefont {Lai}},\ }\bibfield  {title} {\enquote
  {\bibinfo {title} {Linearly polarized remote-edge luminescence in gase
  nanoslabs},}\ }\href {http://arxiv.org/abs/1502.06070} {\bibfield  {journal}
  {\bibinfo  {journal} {arXiv:1502.06070}\ } (\bibinfo {year}
  {2015}{\natexlab{b}})}\BibitemShut {NoStop}%
\bibitem [{\citenamefont {Luttinger}\ and\ \citenamefont
  {Kohn}(1955)}]{luttinger1955}%
  \BibitemOpen
  \bibfield  {author} {\bibinfo {author} {\bibfnamefont {J.~M.}\ \bibnamefont
  {Luttinger}}\ and\ \bibinfo {author} {\bibfnamefont {W.}~\bibnamefont
  {Kohn}},\ }\bibfield  {title} {\enquote {\bibinfo {title} {Motion of
  electrons and holes in perturbed periodic fields},}\ }\href {\doibase
  10.1103/PhysRev.97.869} {\bibfield  {journal} {\bibinfo  {journal} {Phys.
  Rev.}\ }\textbf {\bibinfo {volume} {97}},\ \bibinfo {pages} {869} (\bibinfo
  {year} {1955})}\BibitemShut {NoStop}%
\bibitem [{\citenamefont {Kane}(1957)}]{kane1957}%
  \BibitemOpen
  \bibfield  {author} {\bibinfo {author} {\bibfnamefont {E.~O.}\ \bibnamefont
  {Kane}},\ }\bibfield  {title} {\enquote {\bibinfo {title} {Band structure of
  indium antimonide},}\ }\href {\doibase 10.1016/0022-3697(57)90013-6}
  {\bibfield  {journal} {\bibinfo  {journal} {J. Phys. Chem. Solids}\ }\textbf
  {\bibinfo {volume} {1}},\ \bibinfo {pages} {249--261} (\bibinfo {year}
  {1957})}\BibitemShut {NoStop}%
\bibitem [{\citenamefont {Fishman}\ and\ \citenamefont
  {Lampel}(1977)}]{fishman1977}%
  \BibitemOpen
  \bibfield  {author} {\bibinfo {author} {\bibfnamefont {G.}~\bibnamefont
  {Fishman}}\ and\ \bibinfo {author} {\bibfnamefont {G.}~\bibnamefont
  {Lampel}},\ }\bibfield  {title} {\enquote {\bibinfo {title} {Spin relaxation
  of photoelectrons in p-type gallium arsenide},}\ }\href {\doibase
  10.1103/PhysRevB.16.820} {\bibfield  {journal} {\bibinfo  {journal} {Phys.
  Rev. B}\ }\textbf {\bibinfo {volume} {16}},\ \bibinfo {pages} {820} (\bibinfo
  {year} {1977})}\BibitemShut {NoStop}%
\bibitem [{\citenamefont {Zerrouati}\ \emph {et~al.}(1988)\citenamefont
  {Zerrouati}, \citenamefont {Fabre}, \citenamefont {Bacquet}, \citenamefont
  {Bandet}, \citenamefont {Frandon}, \citenamefont {Lampel},\ and\
  \citenamefont {Paget}}]{zerrouati1988}%
  \BibitemOpen
  \bibfield  {author} {\bibinfo {author} {\bibfnamefont {K.}~\bibnamefont
  {Zerrouati}}, \bibinfo {author} {\bibfnamefont {F.}~\bibnamefont {Fabre}},
  \bibinfo {author} {\bibfnamefont {G.}~\bibnamefont {Bacquet}}, \bibinfo
  {author} {\bibfnamefont {J.}~\bibnamefont {Bandet}}, \bibinfo {author}
  {\bibfnamefont {J.}~\bibnamefont {Frandon}}, \bibinfo {author} {\bibfnamefont
  {G.}~\bibnamefont {Lampel}}, \ and\ \bibinfo {author} {\bibfnamefont
  {D.}~\bibnamefont {Paget}},\ }\bibfield  {title} {\enquote {\bibinfo {title}
  {Spin-lattice relaxation in p-type gallium arsenide single crystals},}\
  }\href {\doibase 10.1103/PhysRevB.37.1334} {\bibfield  {journal} {\bibinfo
  {journal} {Phys. Rev. B}\ }\textbf {\bibinfo {volume} {37}},\ \bibinfo
  {pages} {1334--1341} (\bibinfo {year} {1988})}\BibitemShut {NoStop}%
\bibitem [{\citenamefont {Hilton}\ and\ \citenamefont
  {Tang}(2002)}]{hilton2002}%
  \BibitemOpen
  \bibfield  {author} {\bibinfo {author} {\bibfnamefont {D.~J.}\ \bibnamefont
  {Hilton}}\ and\ \bibinfo {author} {\bibfnamefont {C.~L.}\ \bibnamefont
  {Tang}},\ }\bibfield  {title} {\enquote {\bibinfo {title} {Optical
  orientation and femtosecond relaxation of spin-polarized holes in {GaAs}},}\
  }\href {\doibase 10.1103/PhysRevLett.89.146601} {\bibfield  {journal}
  {\bibinfo  {journal} {Phys. Rev. Lett.}\ }\textbf {\bibinfo {volume} {89}},\
  \bibinfo {pages} {146601} (\bibinfo {year} {2002})}\BibitemShut {NoStop}%
\bibitem [{\citenamefont {Yu}\ \emph {et~al.}(2005)\citenamefont {Yu},
  \citenamefont {Krishnamurthy}, \citenamefont {van Schilfgaarde},\ and\
  \citenamefont {Newman}}]{yu2005}%
  \BibitemOpen
  \bibfield  {author} {\bibinfo {author} {\bibfnamefont {Z.~G.}\ \bibnamefont
  {Yu}}, \bibinfo {author} {\bibfnamefont {S.}~\bibnamefont {Krishnamurthy}},
  \bibinfo {author} {\bibfnamefont {Mark}\ \bibnamefont {van Schilfgaarde}}, \
  and\ \bibinfo {author} {\bibfnamefont {N.}~\bibnamefont {Newman}},\
  }\bibfield  {title} {\enquote {\bibinfo {title} {Spin relaxation of electrons
  and holes in zinc-blende semiconductors},}\ }\href {\doibase
  10.1103/PhysRevB.71.245312} {\bibfield  {journal} {\bibinfo  {journal} {Phys.
  Rev. B}\ }\textbf {\bibinfo {volume} {71}},\ \bibinfo {pages} {245312}
  (\bibinfo {year} {2005})}\BibitemShut {NoStop}%
\bibitem [{\citenamefont {Shen}\ and\ \citenamefont {Wu}(2010)}]{shen2010}%
  \BibitemOpen
  \bibfield  {author} {\bibinfo {author} {\bibfnamefont {K.}~\bibnamefont
  {Shen}}\ and\ \bibinfo {author} {\bibfnamefont {M.~W.}\ \bibnamefont {Wu}},\
  }\bibfield  {title} {\enquote {\bibinfo {title} {Hole spin relaxation in
  intrinsic and $p$-type bulk {GaAs}},}\ }\href {\doibase
  10.1103/PhysRevB.82.115205} {\bibfield  {journal} {\bibinfo  {journal} {Phys.
  Rev. B}\ }\textbf {\bibinfo {volume} {82}},\ \bibinfo {pages} {115205}
  (\bibinfo {year} {2010})}\BibitemShut {NoStop}%
\bibitem [{\citenamefont {Amand}\ and\ \citenamefont
  {Marie}(2008)}]{amand2008}%
  \BibitemOpen
  \bibfield  {author} {\bibinfo {author} {\bibfnamefont {T.}~\bibnamefont
  {Amand}}\ and\ \bibinfo {author} {\bibfnamefont {X.}~\bibnamefont {Marie}},\
  }\enquote {\bibinfo {title} {Exciton spin dynamics in semiconductor quantum
  wells},}\ in\ \href@noop {} {\emph {\bibinfo {booktitle} {Spin Physics in
  Semiconductors}}},\ \bibinfo {editor} {edited by\ \bibinfo {editor}
  {\bibfnamefont {M.~I.}\ \bibnamefont {Dyakonov}}}\ (\bibinfo  {publisher}
  {Springer},\ \bibinfo {year} {2008})\ pp.\ \bibinfo {pages}
  {55--89}\BibitemShut {NoStop}%
\bibitem [{\citenamefont {Boross}\ \emph {et~al.}(2013)\citenamefont {Boross},
  \citenamefont {D{\'o}ra}, \citenamefont {Kiss},\ and\ \citenamefont
  {Simon}}]{boross2013}%
  \BibitemOpen
  \bibfield  {author} {\bibinfo {author} {\bibfnamefont {P.}~\bibnamefont
  {Boross}}, \bibinfo {author} {\bibfnamefont {B.}~\bibnamefont {D{\'o}ra}},
  \bibinfo {author} {\bibfnamefont {A.}~\bibnamefont {Kiss}}, \ and\ \bibinfo
  {author} {\bibfnamefont {F.}~\bibnamefont {Simon}},\ }\bibfield  {title}
  {\enquote {\bibinfo {title} {A unified theory of spin-relaxation due to
  spin-orbit coupling in metals and semiconductors},}\ }\href {\doibase
  10.1038/srep03233} {\bibfield  {journal} {\bibinfo  {journal} {Sci. Rep.}\
  }\textbf {\bibinfo {volume} {3}},\ \bibinfo {pages} {3233} (\bibinfo {year}
  {2013})}\BibitemShut {NoStop}%
\bibitem [{\citenamefont {D'yakonov}\ and\ \citenamefont
  {Perel'}(1971)}]{dyakonov1971}%
  \BibitemOpen
  \bibfield  {author} {\bibinfo {author} {\bibfnamefont {M.~I.}\ \bibnamefont
  {D'yakonov}}\ and\ \bibinfo {author} {\bibfnamefont {V.~I.}\ \bibnamefont
  {Perel'}},\ }\bibfield  {title} {\enquote {\bibinfo {title} {Spin orientation
  of electrons associated with the interband absorption of light in
  semiconductors},}\ }\href
  {http://www.jetp.ac.ru/cgi-bin/dn/e_033_05_1053.pdf} {\bibfield  {journal}
  {\bibinfo  {journal} {Sov. Phys. JETP}\ }\textbf {\bibinfo {volume} {33}},\
  \bibinfo {pages} {1053} (\bibinfo {year} {1971})}\BibitemShut {NoStop}%
\bibitem [{\citenamefont {D'yakonov}\ and\ \citenamefont
  {Perel'}(1972)}]{dyakonov1972}%
  \BibitemOpen
  \bibfield  {author} {\bibinfo {author} {\bibfnamefont {M.~I.}\ \bibnamefont
  {D'yakonov}}\ and\ \bibinfo {author} {\bibfnamefont {V.~I.}\ \bibnamefont
  {Perel'}},\ }\bibfield  {title} {\enquote {\bibinfo {title} {Spin relaxation
  of conduction electrons in noncentrosymmetric semiconductors},}\ }\href@noop
  {} {\bibfield  {journal} {\bibinfo  {journal} {Sov. Phys. Solid State}\
  }\textbf {\bibinfo {volume} {13}},\ \bibinfo {pages} {3023--3026} (\bibinfo
  {year} {1972})}\BibitemShut {NoStop}%
\bibitem [{\citenamefont {Elliott}(1954)}]{elliott1954}%
  \BibitemOpen
  \bibfield  {author} {\bibinfo {author} {\bibfnamefont {R.~J.}\ \bibnamefont
  {Elliott}},\ }\bibfield  {title} {\enquote {\bibinfo {title} {Theory of the
  effect of spin-orbit coupling on magnetic resonance in some
  semiconductors},}\ }\href {\doibase 10.1103/PhysRev.96.266} {\bibfield
  {journal} {\bibinfo  {journal} {Phys. Rev.}\ }\textbf {\bibinfo {volume}
  {96}},\ \bibinfo {pages} {266--279} (\bibinfo {year} {1954})}\BibitemShut
  {NoStop}%
\bibitem [{\citenamefont {Yafet}(1963)}]{yafet1963}%
  \BibitemOpen
  \bibfield  {author} {\bibinfo {author} {\bibfnamefont {Y.}~\bibnamefont
  {Yafet}},\ }\enquote {\bibinfo {title} {g factors and spin-lattice relaxation
  of conduction electrons},}\ in\ \href {\doibase
  10.1016/S0081-1947(08)60259-3} {\emph {\bibinfo {booktitle} {Solid State
  Physics}}},\ \bibinfo {series} {Solid State Physics}, Vol.~\bibinfo {volume}
  {14}\ (\bibinfo  {publisher} {Academic Press},\ \bibinfo {year} {1963})\ pp.\
  \bibinfo {pages} {1--98}\BibitemShut {NoStop}%
\bibitem [{\citenamefont {Bir}\ and\ \citenamefont {Pikus}(1973)}]{bir1973}%
  \BibitemOpen
  \bibfield  {author} {\bibinfo {author} {\bibfnamefont {G.~L.}\ \bibnamefont
  {Bir}}\ and\ \bibinfo {author} {\bibfnamefont {G.~E.}\ \bibnamefont
  {Pikus}},\ }\bibfield  {title} {\enquote {\bibinfo {title} {Optical
  orientation of excitons in uniaxial crystals. {L}arge exchange splitting},}\
  }\href@noop {} {\bibfield  {journal} {\bibinfo  {journal} {Sov. Phys. JETP}\
  }\textbf {\bibinfo {volume} {37}},\ \bibinfo {pages} {1116} (\bibinfo {year}
  {1973})}\BibitemShut {NoStop}%
\bibitem [{\citenamefont {Pikus}\ and\ \citenamefont {Bir}(1974)}]{pikus1974}%
  \BibitemOpen
  \bibfield  {author} {\bibinfo {author} {\bibfnamefont {G.~E.}\ \bibnamefont
  {Pikus}}\ and\ \bibinfo {author} {\bibfnamefont {G.~L.}\ \bibnamefont
  {Bir}},\ }\bibfield  {title} {\enquote {\bibinfo {title} {Optical orientation
  of excitons in cubic crystals},}\ }\href@noop {} {\bibfield  {journal}
  {\bibinfo  {journal} {Sov. Phys. JETP}\ }\textbf {\bibinfo {volume} {40}},\
  \bibinfo {pages} {390--395} (\bibinfo {year} {1974})}\BibitemShut {NoStop}%
\bibitem [{\citenamefont {Bir}\ \emph {et~al.}(1975)\citenamefont {Bir},
  \citenamefont {Aronov},\ and\ \citenamefont {Pikus}}]{bir1975}%
  \BibitemOpen
  \bibfield  {author} {\bibinfo {author} {\bibfnamefont {G.~L.}\ \bibnamefont
  {Bir}}, \bibinfo {author} {\bibfnamefont {A.~G.}\ \bibnamefont {Aronov}}, \
  and\ \bibinfo {author} {\bibfnamefont {G.~E.}\ \bibnamefont {Pikus}},\
  }\bibfield  {title} {\enquote {\bibinfo {title} {Spin relaxation of electrons
  due to scattering by holes},}\ }\href@noop {} {\bibfield  {journal} {\bibinfo
   {journal} {Sov. Phys. JETP}\ }\textbf {\bibinfo {volume} {42}},\ \bibinfo
  {pages} {705--712} (\bibinfo {year} {1975})}\BibitemShut {NoStop}%
\bibitem [{\citenamefont {Aronov}\ \emph {et~al.}(1983)\citenamefont {Aronov},
  \citenamefont {Pikus},\ and\ \citenamefont {Titkov}}]{aronov1983}%
  \BibitemOpen
  \bibfield  {author} {\bibinfo {author} {\bibfnamefont {A.~G.}\ \bibnamefont
  {Aronov}}, \bibinfo {author} {\bibfnamefont {G.~E.}\ \bibnamefont {Pikus}}, \
  and\ \bibinfo {author} {\bibfnamefont {A.~N.}\ \bibnamefont {Titkov}},\
  }\bibfield  {title} {\enquote {\bibinfo {title} {Spin relaxation of
  conduction electrons in p-type {Ill-V} compounds},}\ }\href@noop {}
  {\bibfield  {journal} {\bibinfo  {journal} {Sov. Phys. JETP}\ }\textbf
  {\bibinfo {volume} {57}},\ \bibinfo {pages} {680} (\bibinfo {year}
  {1983})}\BibitemShut {NoStop}%
\bibitem [{\citenamefont {N{\"u}sse}\ \emph {et~al.}(1997)\citenamefont
  {N{\"u}sse}, \citenamefont {Haring~Bolivar}, \citenamefont {Kurz},
  \citenamefont {Klimov},\ and\ \citenamefont {Levy}}]{nusse1997}%
  \BibitemOpen
  \bibfield  {author} {\bibinfo {author} {\bibfnamefont {S.}~\bibnamefont
  {N{\"u}sse}}, \bibinfo {author} {\bibfnamefont {P.}~\bibnamefont
  {Haring~Bolivar}}, \bibinfo {author} {\bibfnamefont {H.}~\bibnamefont
  {Kurz}}, \bibinfo {author} {\bibfnamefont {V.}~\bibnamefont {Klimov}}, \ and\
  \bibinfo {author} {\bibfnamefont {F.}~\bibnamefont {Levy}},\ }\bibfield
  {title} {\enquote {\bibinfo {title} {Carrier cooling and exciton formation in
  {GaSe}},}\ }\href {\doibase 10.1103/PhysRevB.56.4578} {\bibfield  {journal}
  {\bibinfo  {journal} {Phys. Rev. B}\ }\textbf {\bibinfo {volume} {56}},\
  \bibinfo {pages} {4578--4583} (\bibinfo {year} {1997})}\BibitemShut {NoStop}%
\bibitem [{\citenamefont {Kuroda}\ \emph {et~al.}(1980)\citenamefont {Kuroda},
  \citenamefont {Munakata},\ and\ \citenamefont {Nishina}}]{kuroda1980}%
  \BibitemOpen
  \bibfield  {author} {\bibinfo {author} {\bibfnamefont {N.}~\bibnamefont
  {Kuroda}}, \bibinfo {author} {\bibfnamefont {I.}~\bibnamefont {Munakata}}, \
  and\ \bibinfo {author} {\bibfnamefont {Y.}~\bibnamefont {Nishina}},\
  }\bibfield  {title} {\enquote {\bibinfo {title} {Exciton transitions from
  spin-orbit split off valence bands in layer compound {InSe}},}\ }\href
  {\doibase 10.1016/0038-1098(80)90753-X} {\bibfield  {journal} {\bibinfo
  {journal} {Solid State Commun.}\ }\textbf {\bibinfo {volume} {33}},\ \bibinfo
  {pages} {687 -- 691} (\bibinfo {year} {1980})}\BibitemShut {NoStop}%
\bibitem [{\citenamefont {Kuroda}\ and\ \citenamefont
  {Nishina}(1981)}]{kuroda1981}%
  \BibitemOpen
  \bibfield  {author} {\bibinfo {author} {\bibfnamefont {N.}~\bibnamefont
  {Kuroda}}\ and\ \bibinfo {author} {\bibfnamefont {Y.}~\bibnamefont
  {Nishina}},\ }\bibfield  {title} {\enquote {\bibinfo {title} {Anisotropies of
  energy bands in {GaSe} and {InSe}},}\ }\href {\doibase
  10.1016/0378-4363(81)90209-6} {\bibfield  {journal} {\bibinfo  {journal}
  {Physica B+C}\ }\textbf {\bibinfo {volume} {105}},\ \bibinfo {pages} {30--34}
  (\bibinfo {year} {1981})}\BibitemShut {NoStop}%
\bibitem [{\citenamefont {Damen}\ \emph {et~al.}(1991)\citenamefont {Damen},
  \citenamefont {Vi\~{n}a}, \citenamefont {Cunningham}, \citenamefont {Shah},\
  and\ \citenamefont {Sham}}]{damen1991}%
  \BibitemOpen
  \bibfield  {author} {\bibinfo {author} {\bibfnamefont {T.~C.}\ \bibnamefont
  {Damen}}, \bibinfo {author} {\bibfnamefont {L.}~\bibnamefont {Vi\~{n}a}},
  \bibinfo {author} {\bibfnamefont {J.~E.}\ \bibnamefont {Cunningham}},
  \bibinfo {author} {\bibfnamefont {J.}~\bibnamefont {Shah}}, \ and\ \bibinfo
  {author} {\bibfnamefont {L.~J.}\ \bibnamefont {Sham}},\ }\bibfield  {title}
  {\enquote {\bibinfo {title} {Subpicosecond spin relaxation dynamics of
  excitons and free carriers in {GaAs} quantum wells},}\ }\href {\doibase
  10.1103/PhysRevLett.67.3432} {\bibfield  {journal} {\bibinfo  {journal}
  {Phys. Rev. Lett.}\ }\textbf {\bibinfo {volume} {67}},\ \bibinfo {pages}
  {3432--3435} (\bibinfo {year} {1991})}\BibitemShut {NoStop}%
\bibitem [{\citenamefont {Sham}(1993)}]{sham1993}%
  \BibitemOpen
  \bibfield  {author} {\bibinfo {author} {\bibfnamefont {L.~J.}\ \bibnamefont
  {Sham}},\ }\bibfield  {title} {\enquote {\bibinfo {title} {Spin relaxation in
  semiconductor quantum wells},}\ }\href {\doibase 10.1088/0953-8984/5/33A/005}
  {\bibfield  {journal} {\bibinfo  {journal} {J. Phys.: Condens. Matter}\
  }\textbf {\bibinfo {volume} {5}},\ \bibinfo {pages} {A51} (\bibinfo {year}
  {1993})}\BibitemShut {NoStop}%
\bibitem [{\citenamefont {Maialle}\ \emph {et~al.}(1993)\citenamefont
  {Maialle}, \citenamefont {de~Andrada~e Silva},\ and\ \citenamefont
  {Sham}}]{maialle1993}%
  \BibitemOpen
  \bibfield  {author} {\bibinfo {author} {\bibfnamefont {M.~Z.}\ \bibnamefont
  {Maialle}}, \bibinfo {author} {\bibfnamefont {E.~A.}\ \bibnamefont
  {de~Andrada~e Silva}}, \ and\ \bibinfo {author} {\bibfnamefont {L.~J.}\
  \bibnamefont {Sham}},\ }\bibfield  {title} {\enquote {\bibinfo {title}
  {Exciton spin dynamics in quantum wells},}\ }\href {\doibase
  10.1103/PhysRevB.47.15776} {\bibfield  {journal} {\bibinfo  {journal} {Phys.
  Rev. B}\ }\textbf {\bibinfo {volume} {47}},\ \bibinfo {pages} {15776--15788}
  (\bibinfo {year} {1993})}\BibitemShut {NoStop}%
\bibitem [{\citenamefont {Vinattieri}\ \emph {et~al.}(1994)\citenamefont
  {Vinattieri}, \citenamefont {Shah}, \citenamefont {Damen}, \citenamefont
  {Kim}, \citenamefont {Pfeiffer}, \citenamefont {Maialle},\ and\ \citenamefont
  {Sham}}]{vinattieri1994}%
  \BibitemOpen
  \bibfield  {author} {\bibinfo {author} {\bibfnamefont {A.}~\bibnamefont
  {Vinattieri}}, \bibinfo {author} {\bibfnamefont {J.}~\bibnamefont {Shah}},
  \bibinfo {author} {\bibfnamefont {T.~C.}\ \bibnamefont {Damen}}, \bibinfo
  {author} {\bibfnamefont {D.~S.}\ \bibnamefont {Kim}}, \bibinfo {author}
  {\bibfnamefont {L.~N.}\ \bibnamefont {Pfeiffer}}, \bibinfo {author}
  {\bibfnamefont {M.~Z.}\ \bibnamefont {Maialle}}, \ and\ \bibinfo {author}
  {\bibfnamefont {L.~J.}\ \bibnamefont {Sham}},\ }\bibfield  {title} {\enquote
  {\bibinfo {title} {Exciton dynamics in {GaAs} quantum wells under resonant
  excitation},}\ }\href {\doibase 10.1103/PhysRevB.50.10868} {\bibfield
  {journal} {\bibinfo  {journal} {Phys. Rev. B}\ }\textbf {\bibinfo {volume}
  {50}},\ \bibinfo {pages} {10868} (\bibinfo {year} {1994})}\BibitemShut
  {NoStop}%
\bibitem [{\citenamefont {Wang}\ \emph {et~al.}(1995)\citenamefont {Wang},
  \citenamefont {Shah}, \citenamefont {Damen},\ and\ \citenamefont
  {Pfeiffer}}]{wang1995}%
  \BibitemOpen
  \bibfield  {author} {\bibinfo {author} {\bibfnamefont {H.}~\bibnamefont
  {Wang}}, \bibinfo {author} {\bibfnamefont {J.}~\bibnamefont {Shah}}, \bibinfo
  {author} {\bibfnamefont {T.~C.}\ \bibnamefont {Damen}}, \ and\ \bibinfo
  {author} {\bibfnamefont {L.~N.}\ \bibnamefont {Pfeiffer}},\ }\bibfield
  {title} {\enquote {\bibinfo {title} {Spontaneous emission of excitons in
  {GaAs} quantum wells: the role of momentum scattering},}\ }\href {\doibase
  10.1103/PhysRevLett.74.3065} {\bibfield  {journal} {\bibinfo  {journal}
  {Phys. Rev. Lett.}\ }\textbf {\bibinfo {volume} {74}},\ \bibinfo {pages}
  {3065--3068} (\bibinfo {year} {1995})}\BibitemShut {NoStop}%
\bibitem [{\citenamefont {Mu\~{n}oz}\ \emph {et~al.}(1995)\citenamefont
  {Mu\~{n}oz}, \citenamefont {P\'{e}rez}, \citenamefont {Vi\~{n}a},\ and\
  \citenamefont {Ploog}}]{munoz1995}%
  \BibitemOpen
  \bibfield  {author} {\bibinfo {author} {\bibfnamefont {L.}~\bibnamefont
  {Mu\~{n}oz}}, \bibinfo {author} {\bibfnamefont {E.}~\bibnamefont
  {P\'{e}rez}}, \bibinfo {author} {\bibfnamefont {L.}~\bibnamefont {Vi\~{n}a}},
  \ and\ \bibinfo {author} {\bibfnamefont {K.}~\bibnamefont {Ploog}},\
  }\bibfield  {title} {\enquote {\bibinfo {title} {Spin relaxation in intrinsic
  {GaAs} quantum wells: Influence of excitonic localization},}\ }\href
  {\doibase 10.1103/PhysRevB.51.4247} {\bibfield  {journal} {\bibinfo
  {journal} {Phys. Rev. B}\ }\textbf {\bibinfo {volume} {51}},\ \bibinfo
  {pages} {4247--4257} (\bibinfo {year} {1995})}\BibitemShut {NoStop}%
\bibitem [{\citenamefont {Baylac}\ \emph {et~al.}(1995)\citenamefont {Baylac},
  \citenamefont {Amand}, \citenamefont {Brousseau}, \citenamefont {Marie},
  \citenamefont {Dareys}, \citenamefont {Bacquet}, \citenamefont {Barrau},\
  and\ \citenamefont {Planel}}]{baylac1995}%
  \BibitemOpen
  \bibfield  {author} {\bibinfo {author} {\bibfnamefont {B.}~\bibnamefont
  {Baylac}}, \bibinfo {author} {\bibfnamefont {T.}~\bibnamefont {Amand}},
  \bibinfo {author} {\bibfnamefont {M.}~\bibnamefont {Brousseau}}, \bibinfo
  {author} {\bibfnamefont {X.}~\bibnamefont {Marie}}, \bibinfo {author}
  {\bibfnamefont {B.}~\bibnamefont {Dareys}}, \bibinfo {author} {\bibfnamefont
  {G.}~\bibnamefont {Bacquet}}, \bibinfo {author} {\bibfnamefont
  {J.}~\bibnamefont {Barrau}}, \ and\ \bibinfo {author} {\bibfnamefont
  {R.}~\bibnamefont {Planel}},\ }\bibfield  {title} {\enquote {\bibinfo {title}
  {Exciton spin relaxation in the {2D} dense excitonic phase: the role of
  exchange interaction},}\ }\href {\doibase 10.1088/0268-1242/10/3/010}
  {\bibfield  {journal} {\bibinfo  {journal} {Semicon. Sci. Technol.}\ }\textbf
  {\bibinfo {volume} {10}},\ \bibinfo {pages} {295} (\bibinfo {year}
  {1995})}\BibitemShut {NoStop}%
\bibitem [{\citenamefont {Ivchenko}(1995)}]{ivchenko1995}%
  \BibitemOpen
  \bibfield  {author} {\bibinfo {author} {\bibfnamefont {E.~L.}\ \bibnamefont
  {Ivchenko}},\ }\bibfield  {title} {\enquote {\bibinfo {title} {Spectroscopy
  of spin-polarized excitons in semiconductors},}\ }\href {\doibase
  10.1351/pac199567030463} {\bibfield  {journal} {\bibinfo  {journal} {Pure and
  Appl. Chem.}\ }\textbf {\bibinfo {volume} {67}},\ \bibinfo {pages} {463--463}
  (\bibinfo {year} {1995})}\BibitemShut {NoStop}%
\bibitem [{\citenamefont {Amand}\ \emph {et~al.}(1997)\citenamefont {Amand},
  \citenamefont {Robart}, \citenamefont {Marie}, \citenamefont {Brousseau},
  \citenamefont {Le~Jeune},\ and\ \citenamefont {Barrau}}]{amand1997}%
  \BibitemOpen
  \bibfield  {author} {\bibinfo {author} {\bibfnamefont {T.}~\bibnamefont
  {Amand}}, \bibinfo {author} {\bibfnamefont {D.}~\bibnamefont {Robart}},
  \bibinfo {author} {\bibfnamefont {X.}~\bibnamefont {Marie}}, \bibinfo
  {author} {\bibfnamefont {M.}~\bibnamefont {Brousseau}}, \bibinfo {author}
  {\bibfnamefont {P.}~\bibnamefont {Le~Jeune}}, \ and\ \bibinfo {author}
  {\bibfnamefont {J.}~\bibnamefont {Barrau}},\ }\bibfield  {title} {\enquote
  {\bibinfo {title} {Spin relaxation in polarized interacting exciton gas in
  quantum wells},}\ }\href {\doibase 10.1103/PhysRevB.55.9880} {\bibfield
  {journal} {\bibinfo  {journal} {Phys. Rev. B}\ }\textbf {\bibinfo {volume}
  {55}},\ \bibinfo {pages} {9880} (\bibinfo {year} {1997})}\BibitemShut
  {NoStop}%
\bibitem [{\citenamefont {Le~Jeune}\ \emph {et~al.}(1998)\citenamefont
  {Le~Jeune}, \citenamefont {Marie}, \citenamefont {Amand}, \citenamefont
  {Romstad}, \citenamefont {Perez}, \citenamefont {Barrau},\ and\ \citenamefont
  {Brousseau}}]{le-jeune1998}%
  \BibitemOpen
  \bibfield  {author} {\bibinfo {author} {\bibfnamefont {P.}~\bibnamefont
  {Le~Jeune}}, \bibinfo {author} {\bibfnamefont {X.}~\bibnamefont {Marie}},
  \bibinfo {author} {\bibfnamefont {T.}~\bibnamefont {Amand}}, \bibinfo
  {author} {\bibfnamefont {F.}~\bibnamefont {Romstad}}, \bibinfo {author}
  {\bibfnamefont {F.}~\bibnamefont {Perez}}, \bibinfo {author} {\bibfnamefont
  {J.}~\bibnamefont {Barrau}}, \ and\ \bibinfo {author} {\bibfnamefont
  {M.}~\bibnamefont {Brousseau}},\ }\bibfield  {title} {\enquote {\bibinfo
  {title} {Spin-dependent exciton-exciton interactions in quantum wells},}\
  }\href {\doibase 10.1103/PhysRevB.58.4853} {\bibfield  {journal} {\bibinfo
  {journal} {Phys. Rev. B}\ }\textbf {\bibinfo {volume} {58}},\ \bibinfo
  {pages} {4853--4859} (\bibinfo {year} {1998})}\BibitemShut {NoStop}%
\bibitem [{\citenamefont {Vi\~{n}a}(1999)}]{vina1999}%
  \BibitemOpen
  \bibfield  {author} {\bibinfo {author} {\bibfnamefont {L.}~\bibnamefont
  {Vi\~{n}a}},\ }\bibfield  {title} {\enquote {\bibinfo {title} {Spin
  relaxation in low-dimensional systems},}\ }\href {\doibase
  10.1088/0953-8984/11/31/304} {\bibfield  {journal} {\bibinfo  {journal} {J.
  Phys.: Condens. Matter}\ }\textbf {\bibinfo {volume} {11}},\ \bibinfo {pages}
  {5929} (\bibinfo {year} {1999})}\BibitemShut {NoStop}%
\bibitem [{\citenamefont {Feldmann}\ \emph {et~al.}(1987)\citenamefont
  {Feldmann}, \citenamefont {Peter}, \citenamefont {G\"{o}bel}, \citenamefont
  {Dawson}, \citenamefont {Moore}, \citenamefont {Foxon},\ and\ \citenamefont
  {Elliott}}]{feldmann1987}%
  \BibitemOpen
  \bibfield  {author} {\bibinfo {author} {\bibfnamefont {J.}~\bibnamefont
  {Feldmann}}, \bibinfo {author} {\bibfnamefont {G.}~\bibnamefont {Peter}},
  \bibinfo {author} {\bibfnamefont {E.~O.}\ \bibnamefont {G\"{o}bel}}, \bibinfo
  {author} {\bibfnamefont {P.}~\bibnamefont {Dawson}}, \bibinfo {author}
  {\bibfnamefont {K.}~\bibnamefont {Moore}}, \bibinfo {author} {\bibfnamefont
  {C.}~\bibnamefont {Foxon}}, \ and\ \bibinfo {author} {\bibfnamefont {R.~J.}\
  \bibnamefont {Elliott}},\ }\bibfield  {title} {\enquote {\bibinfo {title}
  {Linewidth dependence of radiative exciton lifetimes in quantum wells},}\
  }\href {\doibase 10.1103/PhysRevLett.60.243.4} {\bibfield  {journal}
  {\bibinfo  {journal} {Phys. Rev. Lett.}\ }\textbf {\bibinfo {volume} {59}},\
  \bibinfo {pages} {2337--2340} (\bibinfo {year} {1987})}\BibitemShut {NoStop}%
\bibitem [{\citenamefont {Srinivas}\ \emph {et~al.}(1992)\citenamefont
  {Srinivas}, \citenamefont {Hryniewicz}, \citenamefont {Chen},\ and\
  \citenamefont {Wood}}]{srinivas1992}%
  \BibitemOpen
  \bibfield  {author} {\bibinfo {author} {\bibfnamefont {V.}~\bibnamefont
  {Srinivas}}, \bibinfo {author} {\bibfnamefont {J.}~\bibnamefont
  {Hryniewicz}}, \bibinfo {author} {\bibfnamefont {Y.~J.}\ \bibnamefont
  {Chen}}, \ and\ \bibinfo {author} {\bibfnamefont {C.~E.~C.}\ \bibnamefont
  {Wood}},\ }\bibfield  {title} {\enquote {\bibinfo {title} {Intrinsic
  linewidths and radiative lifetimes of free excitons in {GaAs} quantum
  wells},}\ }\href {\doibase 10.1103/PhysRevB.46.10193} {\bibfield  {journal}
  {\bibinfo  {journal} {Phys. Rev. B}\ }\textbf {\bibinfo {volume} {46}},\
  \bibinfo {pages} {10193--10196} (\bibinfo {year} {1992})}\BibitemShut
  {NoStop}%
\bibitem [{\citenamefont {Citrin}(1993)}]{citrin1993}%
  \BibitemOpen
  \bibfield  {author} {\bibinfo {author} {\bibfnamefont {D.~S.}\ \bibnamefont
  {Citrin}},\ }\bibfield  {title} {\enquote {\bibinfo {title} {Radiative
  lifetimes of excitons in quantum wells: Localization and phase-coherence
  effects},}\ }\href {\doibase 10.1103/PhysRevB.47.3832} {\bibfield  {journal}
  {\bibinfo  {journal} {Phys. Rev. B}\ }\textbf {\bibinfo {volume} {47}},\
  \bibinfo {pages} {3832--3841} (\bibinfo {year} {1993})}\BibitemShut {NoStop}%
\bibitem [{\citenamefont {Aaviksoo}(1991)}]{aaviksoo1991}%
  \BibitemOpen
  \bibfield  {author} {\bibinfo {author} {\bibfnamefont {J.}~\bibnamefont
  {Aaviksoo}},\ }\bibfield  {title} {\enquote {\bibinfo {title} {Time-resolved
  studies of excitonic polaritons},}\ }\href {\doibase
  10.1016/0022-2313(91)90076-8} {\bibfield  {journal} {\bibinfo  {journal} {J.
  Lumin.}\ }\textbf {\bibinfo {volume} {48 \& 49}},\ \bibinfo {pages} {57--66}
  (\bibinfo {year} {1991})}\BibitemShut {NoStop}%
\end{thebibliography}%
\end{document}